%% file: main.tex
\documentclass[conference]{IEEEtran}

\ifCLASSINFOpdf
\else
\fi

\usepackage[available,functional,reproduced]{ndssbadges}
\usepackage{hyperref}
\usepackage[export]{adjustbox}
\usepackage[normalem]{ulem}
\usepackage{multirow}
\usepackage{pifont}
\usepackage{listings}
\usepackage{mathpartir}
\usepackage{xspace}
\usepackage{tikz}
\usepackage{graphicx}
\usepackage{siunitx}
\usepackage{mdframed}
\usepackage{mathtools}
\usepackage[nounderscore]{syntax}
\usepackage{subcaption}
\usepackage[linesnumbered,ruled,vlined]{algorithm2e}
\usepackage{balance}
\usepackage{mathtools}
\usepackage{mathpartir}
\usepackage{amsthm}
\usepackage[export]{adjustbox}
\usepackage{xcolor,colortbl}
\usepackage{multirow}
\usepackage{amsfonts}
\usepackage{booktabs}
\usepackage{stmaryrd}
\usepackage{amssymb}
\newcommand{\langkw}[1]{\textbf{\color{blue}#1}}

\usepackage{multicol}
\usepackage[most]{tcolorbox}

\usepackage{tikz}
\usepackage{amsmath}

\lstdefinestyle{customc}{
  belowcaptionskip=1\baselineskip,
  breaklines=true,
  numbers=left,
  xleftmargin=2.5em,
  language=C,
  showstringspaces=false,
  basicstyle=\footnotesize\ttfamily,
  keywordstyle=\bfseries\color{black},
  commentstyle=\itshape\color{gray},
  identifierstyle=\color{black},
  stringstyle=\color{orange},
  morekeywords={size_t},
  moredelim=**[is][\color{red}]{@}{@},
  moredelim=**[is][\color{green!40!black}]{@@}{@@},
}

\usepackage{tikz}
\usepackage{xcolor}
\newcommand*\circled[1]{\tikz[baseline=(char.base)]{
            \node[shape=circle,fill,inner sep=1.2pt] (char) {\textcolor{white}{#1}};}}

\DeclareMathOperator*{\argmax}{argmax}

\SetKwProg{Func}{Function}{}{}
\SetAlFnt{\small\rmfamily}
\SetKw{Continue}{continue}
\SetKw{Break}{break}
\SetKw{Case}{case}
\SetKw{And}{and}
\SetKw{Or}{or}
\SetKw{Not}{not}
\SetKw{In}{in}
\SetKw{Is}{is}
\SetKw{Null}{NULL}
\SetKwInput{Input}{Input}
\SetKwInOut{Output}{Output}

\newcommand{\revise}[1]{{\color{black}{#1}}}

\newcommand{\toolname}{{\sc GenNm}\xspace}

\definecolor{Gray}{gray}{0.935}
\newcolumntype{g}{>{\columncolor{Gray}}c}
\newcolumntype{G}{>{\columncolor{Gray}}r}

\newtcolorbox{myquote}[1][]{%
    colback=black!5,
    colframe=black!5,
    notitle,
    sharp corners,
    borderline west={2pt}{0pt}{red!80!black},
    enhanced,
    breakable,
    }

\newenvironment{reviewer-comment}{}{}
\tcbuselibrary{skins}
\tcolorboxenvironment{reviewer-comment}{empty,
  left = 1em, top = 1ex, bottom = 1ex,
  borderline west = {2pt} {0pt} {black!20},
}
\ExplSyntaxOn
\NewDocumentEnvironment {response} { +m O{black!20} } {
  \IfValueT {#1} {
    \begin{reviewer-comment}
      \setlength\parindent{2em}
      \noindent
      \ttfamily #1
    \end{reviewer-comment}
  }
  \par\noindent\ignorespaces
} { \par }
\ExplSyntaxOff

\begin{document}
\title{Unleashing the Power of Generative Model\\ in Recovering Variable Names from Stripped Binary}

\author{\IEEEauthorblockN{Xiangzhe Xu}
\IEEEauthorblockA{Purdue University\\
xzx@purdue.edu}
\and
\IEEEauthorblockN{Zhuo Zhang}
\IEEEauthorblockA{Purdue University\\ zhan3299@purdue.edu}
\and
\IEEEauthorblockN{Zian Su}
\IEEEauthorblockA{Purdue University\\ su284@purdue.edu}
\and
\IEEEauthorblockN{Ziyang Huang}
\IEEEauthorblockA{Purdue University\\ huan1562@purdue.edu}
\and
\IEEEauthorblockN{Shiwei Feng}
\IEEEauthorblockA{Purdue University\\ feng292@purdue.edu}
\and
\IEEEauthorblockN{Yapeng Ye}
\IEEEauthorblockA{Purdue University\\ ye203@purdue.edu}
\and
\IEEEauthorblockN{Nan Jiang}
\IEEEauthorblockA{Purdue University\\ jiang719@purdue.edu}
\and
\IEEEauthorblockN{Danning Xie}
\IEEEauthorblockA{Purdue University\\ xie342@purdue.edu}
\and
\IEEEauthorblockN{Siyuan Cheng}
\IEEEauthorblockA{Purdue University\\ cheng535@purdue.edu}
\and
\IEEEauthorblockN{Lin Tan}
\IEEEauthorblockA{Purdue University\\ lintan@purdue.edu}
\and
\IEEEauthorblockN{Xiangyu Zhang}
\IEEEauthorblockA{Purdue University\\ xyzhang@cs.purdue.edu}
}

\IEEEoverridecommandlockouts
\makeatletter\def\@IEEEpubidpullup{6.5\baselineskip}\makeatother
\IEEEpubid{\parbox{\columnwidth}{
    Network and Distributed System Security (NDSS) Symposium 2025\\
    24-28 February 2025, San Diego, CA, USA\\
    ISBN 979-8-9894372-8-3\\
    https://dx.doi.org/10.14722/ndss.2025.240276\\
    www.ndss-symposium.org
}
\hspace{\columnsep}\makebox[\columnwidth]{}}

\maketitle

\begin{abstract}
Decompilation aims to recover the source code form of a binary executable.
It has many security applications, such as malware analysis, vulnerability detection, and code hardening. A prominent challenge in decompilation is to recover variable names. 
We propose a novel technique that leverages the strengths of generative models while 
mitigating model biases.
We build a prototype, \toolname, from pre-trained generative models CodeGemma-2B, CodeLlama-7B, and CodeLlama-34B.
We fine-tune \toolname on decompiled functions and teach models to leverage contextual information.
\toolname includes names from callers and callees while querying a function, providing rich contextual information within the model's input token limitation. 
We mitigate model biases by aligning the output distribution of models with symbol preferences of developers.
Our results show that \toolname improves the state-of-the-art name recovery precision by 5.6--11.4 percentage points on two commonly used datasets and improves the state-of-the-art by 32\% (from 17.3\% to 22.8\%) in the most challenging setup where ground-truth variable names are not seen in the training dataset.
\end{abstract}

\input{introduction}

\input{moti}

\input{method}

\input{evaluation}

\input{related-and-conclusion}

\section*{Acknowledgment}

We thank the anonymous reviewers for their valuable comments and suggestions. We are grateful to the Center for AI Safety for providing computational resources. This research
was supported in part by DARPA VSPELLS - HR001120S0058, IARPA TrojAI W911NF-19-S-0012,
NSF 1901242 and 1910300, ONR N000141712045, N000141410468 and N000141712947. Any opinions,
findings, and conclusions in this paper are those of the authors only and do not necessarily reflect the views of our sponsors.

\bibliographystyle{IEEEtranS}
\bibliography{IEEEabrv,reference}

\clearpage
\newpage

\appendix

\input{appendix}

\input{artifact}

\end{document}

%% file: introduction.tex
\section{Introduction}
\label{sec:intro}
Deployed software often has the form of binary executable. 
Understanding these prevalent binaries is essential for various security and safety aspects of software, including conducting security assessments of contemporary devices such as %
home automation systems~\cite{ioannis2023firmsolo,gustafson2023shimware} and autopilot technology~\cite{tesla2023zq}, 
maintaining and hardening legacy software~\cite{controlflow2015carlini,martin2010dynamically,carbone2009mapping}, 
detecting vulnerabilities in commercial-off-the-shelf software~\cite{xu2017spain,li2017semhunt,xu2023pem,su2024codeart,xu2023improving}, 
and analyzing malware that threatens our daily lives~\cite{xu2014autoprobe, spensky2016phi,simone23malware}.
A significant challenge, however, is presented by the fact that most of these binaries are shipped without source code, making them extremely difficult to comprehend.
To bridge this gap, reverse engineering techniques have emerged to recover source-code-level details.
Over the past decade, techniques like disassembly~\cite{bauman2018superset,peixda,miller2019probabilistic,altinay2020binrec,ye2023d}, function boundary identification~\cite{alves2019function}, type inference~\cite{lin2010rewards,shoshitaishvili2016state,lee2011tie,slowinska2011howard,zhang23osprey,pei2021stateformer}, recovery of high-level abstractions~\cite{schwartz2018using,jin2023binary,xu2023lmpa}, code structure~\cite{basque2024ahoy}, and data structures~\cite{schwartz2018using} have advanced significantly.

Despite these successes, a crucial step of reverse-engineering pipelines, namely recovering variable names, remains inadequately addressed. 
Recovering names from fully stripped binary programs entails an in-depth understanding of both \textit{machine semantics}, concerning data-flow and control-flow,
and \textit{descriptive semantics}, reflecting how the code is understood by human developers. 
This duality poses a significant challenge for conventional analysis methods~\cite{brumley2013native,yakdan2015no} that primarily focus on machine semantics, resulting in the failure of recovering meaningful variable names. 

Recent work in renaming variables benefits from advances in machine learning models~\cite{dire,dirty,varbert}. They formulate the problem as a classification task (a.k.a., a closed-vocabulary sequence labeling task).
In the training stage, a model learns a set of variable names, i.e., the vocabulary. In the inference stage, it takes as input a decompiled function, and predicts the name for each variable by picking one from the vocabulary. Although such methods have achieved good results, their generality is limited due to the following reasons.
(1) A classification model can only predict names within its training vocabulary. (2) Variable name distributions are largely biased, and it is challenging to train a good classifier on biased distributions~\cite{he2009learning,huang2016learning}.
(3) Existing methods 
process one function at a time, due to models' input size limits or model capacity,  missing important contextual information.

In particular, a classification model can only select names from the training vocabulary. It cannot ``invent'' new names based on program contexts. Consequently, the state-of-the-art model achieves a precision of less than 10\% for variables whose ground-truth names are not in the training dataset (see Section~\ref{sec:perf-general}). 
Moreover, distributions of variable names are \textbf{\textit{biased}}. For example, in datasets constructed from GitHub repositories~\cite{dirty} or Linux packages~\cite{varbert}, more than 50\% of names appear less than 2 times, whereas 0.1\% of names appear over 1,000 times. A typical classification loss that
maximizes the probability of 
selecting
the ground-truth names
would undesirably emphasize the frequent names.
As a result, the performance of  classification model degrades by 57.5\% (from 31.8\% to 13.5\%) for names rarely present in the training dataset (see Section~\ref{sec:perf-general}).
Finally, most existing models used in reverse engineering have a limited input window
and hence analyze individual decompiled functions
independently,
missing important information in the calling context.

To address the challenges, we propose \toolname as a systematic solution to recovering names from fully stripped binaries.
\revise{
\toolname leverages a generative code language model fine-tuned with \textit{contextual information} and \textit{symbol preferences}.
It composes variable names from tokens, and thus has better chance to generalize to rarely present or even unseen names. Intuitively, a human developer seldom considers complex variable names as an atomic word. Instead, she understands it as a composition of several keywords. For example, a rare name {\tt ip\_hdrlen} is composed from three frequent sub-words: {\tt ip}, {\tt header}, and {\tt length}. The generative nature of model
thus naturally aligns with the cognitive model of a human developer. Our evaluation results in Section~\ref{sec:perf-general} show that more than 95\% rare names are composed from commonly appeared tokens.

To leverage the power of a generative model, a typical fine-tuning technique such as that in DIRE~\cite{dire} and ReSym~\cite{resym} simply guides a model to predict names from individual binary functions. Consequently, the trained model has limited knowledge about how to leverage contextual information. Nevertheless, information from the calling context plays a crucial role in understanding binary programs, as the absence of meaningful function names hinders the model's ability to deduce the semantics of calling contexts.
We thus propose a novel context-aware fine-tuning paradigm that teaches a model to reason a binary function with both the function body and the information from its calling contexts. The model thus can learn the relation between names of local variables and names in the calling contexts.

Moreover, we formulate the implicit biases in the training dataset as misalignment between data frequencies and \textit{symbol preference} of developers.
Symbol preference denotes that a developer prefers one name over another in a specific program context. For example, in the context of network programming, a developer typically names the data sent over network as {\tt packet} instead of {\tt array}, although {\tt array} may be a more frequent name in the whole dataset. We therefore propose an additional training stage named \textit{SYMbol Preference Optimization}~(SymPO) to explicitly guide our model to pick the preferred name over the sub-optimal ones under given program contexts.

We design the inference stage of \toolname as an iterative process. It is inspired by recent studies on human reverse engineering~\cite{mantovani2022re,burk2022decomperson}, which emphasize the practice of inspecting all functions in a breadth-first manner, followed by iterative refinement.  Initially, \toolname generates variable names for each function based on local context (i.e., the function body). Next, we traverse the program call graph to propagate contextual information from each function to its caller and callee functions, aligning with the training objective. Inspired by previous work~\cite{resym}, we leverage program analysis and majority voting to pick the final name across different iterations.
}

We summarize our contributions as follows:
\begin{itemize}%
\item \revise{We propose a novel context-aware fine-tuning paradigm that teaches a model to leverage contextual information when reasoning a decompiled function.}
\item We encode the symbol preference for variable names in the training pipeline, guiding the model to select names relevant to program context with higher probabilities.

\item \revise{We design an iterative inference process aligned with the way human reverse engineers leverage contextual information.}

\item We develop a prototype \toolname(Unleashing the Power of \underline{Gen}erative Model in Recovering Variable \underline{N}a\underline{m}es from Stripped Binary). \revise{\toolname improves the name recovery accuracy over the state-of-the-art techniques~\cite{varbert,resym} by 5.6--11.4 percentage points on two commonly used datasets with the most challenging setup. On challenging cases where the ground-truth names are not seen during training, \toolname improves over the state-of-the-art techniques~\cite{varbert,resym} by 168\%~(8.5\% versus 22.8\%) and 32\%~(17.3\% versus 22.8\%), respectively.}
\end{itemize}

%% file: moti.tex
\section{Motivation and Overview}

We use a motivating example to illustrate the limitations of the state-of-the-art technique for renaming decompiled variables. Following this, we demonstrate our method.

\subsection{Motivating Example}

\begin{figure*}
\centering
    \begin{subfigure}[t]{.38\linewidth}
    \centering
        \includegraphics[width=\linewidth]{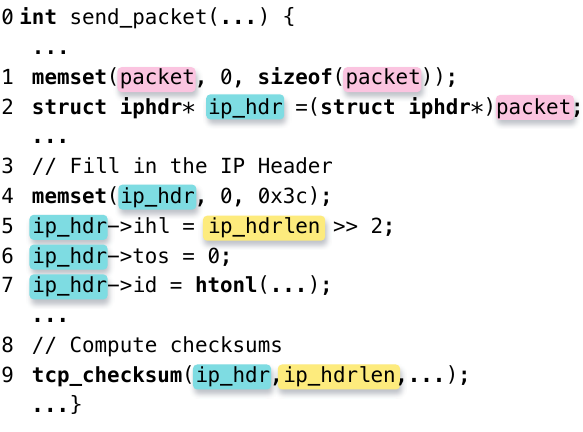}
        \caption{Source code}
        \label{fig:moti:code-src}
    \end{subfigure}    
    \hspace{.02\textwidth}
    \begin{subfigure}[t]{.3\linewidth}
    \centering
        \includegraphics[width=\linewidth]{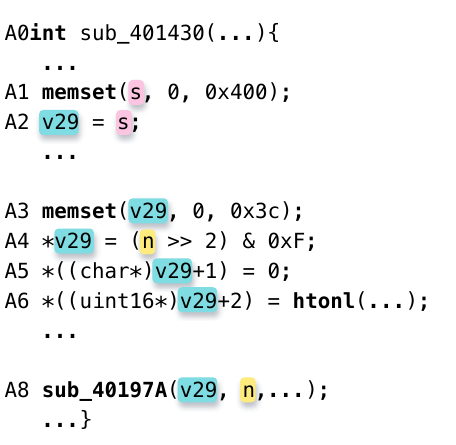}
        \caption{Decompiled code}
        \label{fig:moti:code-dec}
    \end{subfigure}    
    ~\hspace{.02\textwidth}
    \begin{subfigure}[t]{.22\linewidth}
    \centering
        \includegraphics[width=.9\linewidth]{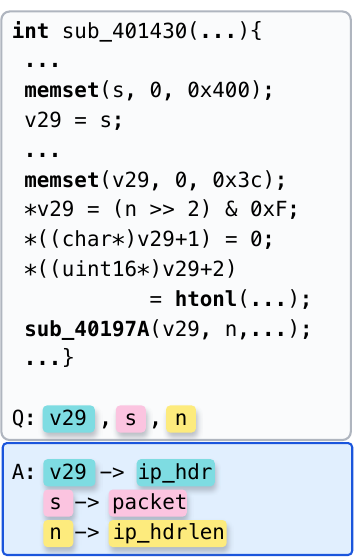}
        \caption{Input to fine-tune \toolname}
        \label{fig:moti:code-input}
    \end{subfigure}    
    \caption{Code snippets for the motivating example. Corresponding variables are highlighted with same colors.}
    \vspace{-10pt}
    \label{fig:moti-code}
\end{figure*}

Fig.~\ref{fig:moti-code}a shows our motivating example, which is adapted from the function {\tt send\_packet()} in an exploit for CVE-2018-4407~\cite{cve20184407}.
The function sends a TCP packet that triggers a buffer-overflow vulnerability. The code snippet in Fig.~\ref{fig:moti-code}a illustrates the logic to initialize the IP packet header ({\tt ip\_hdr}, lines 4--7) and compute the checksum for the TCP packet (line~9). 

We show the corresponding decompiled code in Fig.~\ref{fig:moti-code}b. Each line in the decompiled code is aligned with the related source code line, and the corresponding variables are highlighted with the same colors. For variables {\tt s} and {\tt n} in the decompiled code, the decompiler (IDA-Pro~\cite{idapro} in this case) synthesizes their names based on calls to library functions and the types of the two variables. Although synthesized names %
may help understanding 
(e.g., {\tt n} may indicate the length of some buffer),
they can hardly reflect the context and are hence much less informative than the original symbol names. For example, the source code variable related to {\tt n} is {\tt ip\_hdrlen}, which denotes the length of the IP packet header. The synthesized name {\tt n} fails to reflect this information.
Similarly, the variable name {\tt v29} is simply a placeholder name that is not meaningful. 
In both cases, the variable names in the decompiled code cannot reflect similar semantics to their source code counterparts.

The goal of variable name recovery is hence to generate meaningful names for variables with placeholder names or names synthesized from library functions. 

\subsection{Challenges and Limitations of State-of-the-Art}
\label{sec:moti:limit}

\begin{figure}[t]
\centering
    \includegraphics[width=.85\linewidth]{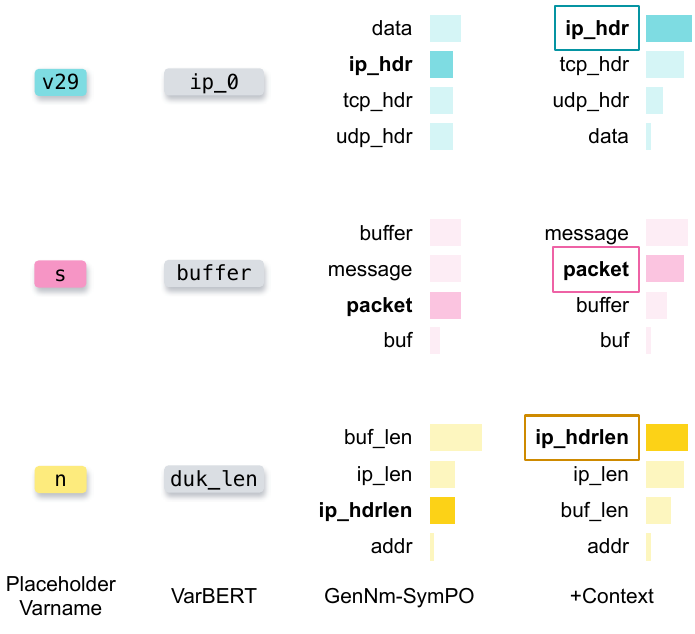}
    \caption{Name selections for baseline (VarBERT) and name distributions for the predictions of \toolname. 
    Each column denotes the predictions of a technique. VarBERT denotes the baseline model,
    \textit{GenNm-SymPO} denotes the \toolname model after fine-tuning and symbol preference optimization. \textit{+Context} denotes the model is used with the contextual information propagated along the call graph. 
    Blue, pink, and yellow colors denote predictions for {\tt v29}, {\tt s}, and {\tt n}. Names are ranked by their probability where a longer bar denotes a higher probability.
    Names highlighted with \textbf{bold fonts} are names similar or equal to ground-truth names. Names with \textit{outlines} 
    are those selected by the name validation algorithm.
    }
    \label{fig:moti:tech-preds}
\end{figure}

The state-of-the-art technique VarBERT~\cite{varbert} leverages the Transformer model~\cite{transformer} to recover variable names. The model takes as input a decompiled function and predicts a name for each variable. The problem is formulated as a classification task. A set of variable names is first collected from the training data, noted as the vocabulary. A model is trained to select a name from the vocabulary for each variable in the decompiled function. We show with the motivating example three major challenges in recovering variables from stripped binaries and thus discuss the limitations of 
state-of-the-art. The predictions of VarBERT for the motivating example are shown in the second column of Fig.~\ref{fig:moti:tech-preds}.

\noindent
{\bf Challenge 1: Cannot predict names not in the vocabulary.}
A classification model can only select names from those seen during training (i.e., the vocabulary). It cannot compose new names based on program contexts. %
In Section~\ref{sec:perf-general},
we will show that 
16\% of the variables in our test dataset have ground-truth names not seen in the training dataset~\footnote{Our dataset is derived from a high-quality dataset VarCorpus, with a train-test split ratio of 9:1. See Section~\ref{sec:setup} for details.}. As a result, 
VarBERT achieves only 
8.5\% precision on those variables. In our motivating example, the ground-truth name for variable {\tt n} (at line A4 in Fig.~\ref{fig:moti-code}b) is {\tt ip\_hdrlen}, which indicates the length of an IP packet header. However, the name never occurs in the training dataset. 
Thus, VarBERT mistakenly predicts the name of {\tt n} as {\tt duk\_len}, where {\tt duk} is an irrelevant program in the training dataset. 
For unseen names, \toolname %
outperforms VarBERT by 168\%, i.e., 8.5\% versus 22.8\%~(details in Section~\ref{sec:perf-general}).

\begin{figure}[t]
    \centering
    \includegraphics[width=.75\linewidth]{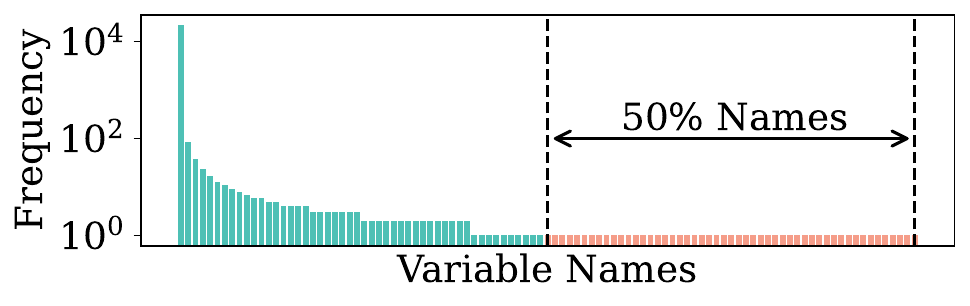}
    \caption{Distribution of name frequencies. More than 50\%~variable names (in orange) appear only once in the training dataset.}
    \label{fig:moti-name-freq}
\end{figure}

\noindent
{\bf Challenge 2: Long-tail distribution of variable names makes correct prediction difficult.} The distribution of variable names is imbalanced and has a long tail. For example, Fig.~\ref{fig:moti-name-freq} shows the distribution of names in our dataset in terms of frequency. Observe that the most frequent name appears around 50k times, while 50\% of the names appear only once. 
It is hence challenging to train a classification model from such  data with a significantly biased distribution~\cite{he2009learning,huang2016learning}. 
A typical classification loss used 
in training optimizes the model's probability to predict the ground-truth name for each variable. This training loss undesirably emphasizes the frequent names.
For example, 
the VarBERT model predicts the name of {\tt s} (at lines A1--A2 in Fig.~\ref{fig:moti-code}b) as {\tt buffer}. We analyze the training dataset and find that the variable name {\tt buffer} is passed as the first argument to {\tt memset} for more than 500 times in the training samples. On the other hand, the ground-truth name in this case, {\tt packet}, appears with {\tt memset} for only 25 times in the training data. Therefore, 
the model biases towards the name {\tt buffer} after seeing the variable is used as the first argument to {\tt memset} in the query. \revise{Please see Appendix~\ref{appdx:corr} for a quantified statistical test.}

\noindent
{\bf Challenge 3: Missing contextual information makes prediction difficult.} 
Limited by the input length and the understanding capability of typical classification models~(which are smaller than pre-trained generative models), VarBERT and many other existing works~\cite{dire,dirty}
analyzes only one function at a time. This practice, however, misses important information from the calling context. For example, at line A8 of Fig.~\ref{fig:moti-code}b, a model has no knowledge about the callee function {\tt sub\_40197A} without contextual information. Consequently, it can hardly deduce the semantics of variable {\tt v29}, which is passed as the first argument to {\tt sub\_40197A}. 
VarBERT mistakenly predicts {\tt v29} as {\tt ip\_0}, while the ground-truth name is {\tt ip\_hdr}.

\subsection{Our Method}
\label{sec:moti-our-tech}
\revise{We alleviate the closed vocabulary problem by fine-tuning generative models that can compose unseen names.
To augment a query function with better context, we propagate information of individual functions through the call graph. We design a new context-aware fine-tuning paradigm to teach the generative model how to predict names considering additional contextual information.
To accommodate the generative model to the biased distribution of variable names, we design \textit{symbol preference optimization} that aligns the model with the symbol preference of developers.
}

\noindent
{\bf Solution for Challenge 1: Fine-tuning generative models.}
A generative model can concatenate multiple tokens to construct a variable name and hence has potential advantages over classification-based methods. 
Large language models (LLMs) (e.g., ChatGPT and GPT-4~\cite{instructgpt, openai2023gpt4}) are advanced pre-trained generative models. They demonstrate strong capabilities in understanding both natural language text and source code. However, the distribution of decompiled code is dissimilar to either. 
Our evaluation in Section~\ref{sec:eval:compare-llm} shows that ChatGPT and GPT-4 underperform our model by 11.3 percentage points in terms of precision.

To bridge the gap between the distribution of the pre-training knowledge in a generative code language model and the distribution of decompiled code, we fine-tune a generative model using decompiled code.
An example input used in the fine-tuning stage is shown in Fig.~\ref{fig:moti-code}c, where the grey box contains the query decompiled function and a list of placeholder variable names; the blue box contains the expected response of \toolname, consisting of a map from a placeholder name to a ground-truth variable name. Intuitively, the 
fine-tuning guides the model to generate the expected variable names based on the query function. The last row at the third column~(GenNm-SymPO) of Fig.~\ref{fig:moti:tech-preds} shows that after fine-tuning, \toolname composes the unseen name {\tt ip\_hdrlen} as a top candidate.

\noindent
{\bf Solution for Challenge 2: Symbol preference optimization~(SymPO).}
Similar to a classification model trained only on ground-truth names, a generative model trained only on ground-truth names inherits the biases in the training dataset. Our key insight is that developers' preference over symbol names is implied by the ground-truth names, and the preference can be used to mitigate the biases in the training dataset.
We propose the concept \textit{symbol preference}, denoting that a name is preferred over other names given certain program context.
For example, the variable marked in pink in Fig.~\ref{fig:moti:tech-preds} has the ground-truth name {\tt packet}. That is because {\tt packet} is more relevant to the context of network programming, and is thus more preferable than the highly frequent name {\tt buffer}.

Technically, after training a generative model with the ground-truth names, we use the trained model to perform inference on the {\it training} dataset. We then collect the cases that the model makes mistakes. Intuitively, these counterexamples reflect the misalignment between the model's biases and the symbol preference. We adapt a loss function used in the {\em  direct preference optimization}~(DPO)~\cite{rafailov2024direct} algorithm, guiding the model to select the preferred names over the biased ones. As a result, as shown in the third column~(GenNm-SymPO) of Fig.~\ref{fig:moti:tech-preds}, after SymPO
the preferred name {\tt packet}~(in pink rows) and {\tt ip\_hdr}~(in blue rows) have high probabilities,
comparable to the most frequent names {\tt buffer}~(in pink rows) and {\tt data}~(in blue rows).~\looseness=-1%

\begin{figure}[t]
\centering
    \includegraphics[width=.43\linewidth]{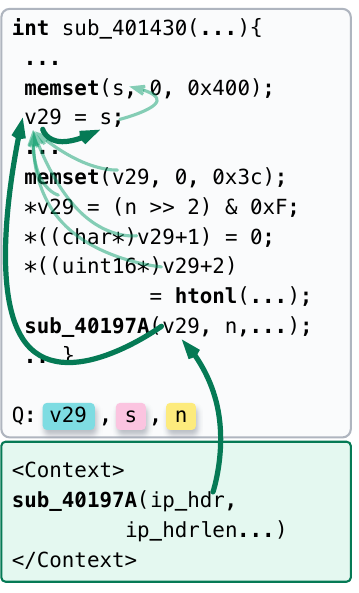}
    \caption{Query prompt to \toolname augmented with the information propagated from the calling context~(the green box). Dataflow used in name validation are indicated by green arrows, with most relevant ones highlighted.%
    }
    \label{fig:moti:tech-input}
\end{figure}    

\revise{
\noindent
{\bf Solution for Challenge 3: Iterative inference and context-aware fine-tuning.}}
Individual decompiled functions have limited contextual information. The local information in a function may not be sufficient for a model to generate correct names. Take {\tt v29} as an example. \toolname generates with high probabilities three similar names: {\tt ip\_hdr}, {\tt tcp\_hdr}, and {\tt udp\_hdr}, as shown at the blue rows of the column~GenNm-SymPO in Fig.~\ref{fig:moti:tech-preds}. However, without the contextual information from the callee function {\tt sub\_40197A}, it is challenging to decide the precise name for {\tt v29}.
A straightforward solution to leverage global contextual information would be including caller and callee code bodies into the query. However, this naive solution incurs a substantially higher cost due to the much larger number of tokens entailed. Moreover, although LLMs have a relatively long context-window length, the performance degrades when the input becomes longer~\cite{hsieh2024ruler}(\revise{detailed discussion is in Appendix~\ref{appdx:long-ctx}}).  Therefore, we use function signatures as summaries for calling contexts. Specifically, we design an iterative inference process. 
We first ask \toolname to generate names based on local information (e.g., the function shown in the grey box of Fig.~\ref{fig:moti-code}c) for individual functions, and then
gather the predicted names along the program call graph, adding contextual information to the queries of individual functions.
For example, the green box in Fig.~\ref{fig:moti:tech-input} shows the context propagated to our motivating example. Note that names {\tt ip\_hdr} and {\tt ip\_hdrlen} in the green box are predicted based on the function body of the callee function {\tt sub\_40197A}~(not shown in the figure).
The last column~(+Context) of Fig.~\ref{fig:moti:tech-preds} shows the output distribution of \toolname when contextual information is introduced to the query. We can see that the model correctly predicts {\tt v29} and {\tt n} with the ground-truth names. 

\revise{A generative model fine-tuned with only the function body and the ground-truth names~(as shown in Fig.~\ref{fig:moti-code}c) may have limited knowledge about how to effectively leverage the contextual information. We therefore design a novel context-aware fine-tuning paradigm, providing contextual information (as shown in the green box in Fig.~\ref{fig:moti:tech-input}) during fine-tuning so that the model can learn the relation between the names of local variables and the names in the calling contexts. According to our experiments in Section~\ref{sec:eval:ablation}, this is the key reason for \toolname's superior performance. 

Finally, to select the best name across multiple inference iterations, we propose a name validation algorithm to select (from top-ranked candidates) the name that is most consistent with the local program context. We propagate names along program data-flow. For example, to select the best name for variable {\tt s}, the data-flow edges highlighted in Fig.~\ref{fig:moti:tech-input} connects it to {\tt v29}, and {\tt v29} is further connected to the first argument {\tt ip\_hdr} of the callee function {\tt sub\_40197A}. They indicate the names of those variables may have semantics relevance with {\tt s}. \toolname calculates the semantics similarity between the names of those variables and the candidate names of {\tt s} (i.e., {\tt message}, {\tt packet}, {\tt buffer}, and {\tt buff}).
It then finds that the name {\tt packet} is the most relevant with the names of the other two variables.
}

%% file: method.tex
\section{Problem Definition}

To facilitate discussion, we formalize the problem as shown in Fig.~\ref{fig:defs}.
We use $id$ to refer to the placeholder names synthesized by the decompiler, and use $name$ to refer to meaningful names. A binary program consists of an id, a list of binary functions, and a call graph. The call graph is a set of edges from caller functions to callee functions.
A decompiled function consists of a function id, the string of decompiled code, and a set of identifiers used in the function.
A name map is associated with a binary program. It takes as input the id of a function, the id of a variable in this function, and returns a meaningful name for the variable. The dataset of binary programs $\mathcal{D}$ has the type of a list of pairs. Each pair consists of a binary program and the corresponding name map containing the ground-truth names.

We transform the decompiled code of a function to a program in a simple language to simplify the discussion. The language definition is shown in the lower part of Fig.~\ref{fig:defs}. The definitions are standard. Note that we omit most types of expressions and only focus on expressions containing an identifier~($id$) and a function call~($id(\mathcal{A})$).

\newcommand{\fieldemph}[1]{\textbf{\textit{#1}}}

\begin{figure}[t]
    \centering
\begin{mdframed}    
\small
    $\mathcal{B}\in$ \texttt{Binary} $\Coloneqq$ $\{
    \textbf{\textit{bid}}: \textit{id},\
    \textbf{\textit{funcs}}:\textit{list}\ \mathcal{F}, \textbf{\textit{cg:}}\textit{set}\ \mathcal{F}\times\mathcal{F}
    \}$
    \\
    $\mathcal{F}\in$ \texttt{Function} $\Coloneqq$ $\{
    \textbf{\textit{fid}}: \textit{id},\ 
    \textbf{\textit{body}}:\textit{str},
     \textbf{\textit{ids}}: \textit{set} \ \textit{id}
     \}$
     \\
    $\mathcal{N}\in$ \texttt{NameMap} $\Coloneqq$ $\textit{id} \rightarrow \textit{id} \rightarrow \textit{str}$
    \\
    $\mathcal{D}\in$ \texttt{Dataset}
    $\Coloneqq$ $\textit{list}\ (\;\mathcal{B}\times\mathcal{N}\;)$
\end{mdframed}
\vspace{-12pt}
\begin{mdframed}
\small
      {
\centering        
  \begin{grammar}
    <\textit{FuncBody}> $\mathcal{F}.\textbf{\textit{body}}$ $\Coloneqq$ $\mathcal{R}; \mathcal{S}$
\quad 
    $\langle\textit{Params}\rangle$ $\mathcal{R}$ $\Coloneqq$ \textit{list} \textit{id}

    <\textit{Expr}> $\mathcal{E}$ $\Coloneqq$ $id$ | $id(\mathcal{A})$ | \textit{Other}    
\quad
    $\langle\textit{Args}\rangle$ $\mathcal{A}$ $\Coloneqq$ \textit{list} $\mathcal{E}$
    
    <\textit{Stmt}> $\mathcal{S}$ $\Coloneqq$ $\mathcal{S}_1;\mathcal{S}_2$ 
    | $\mathcal{E}_0 \coloneqq \mathcal{E}_1$ 
    | \langkw{return} $\mathcal{E}$ 
    | \langkw{while}$(\mathcal{E})\{\mathcal{S}\}$ 
    | \langkw{if}$(\mathcal{E})\{\mathcal{S}_1\}$\langkw{else}$\{\mathcal{S}_2\}$    
  \end{grammar}
      }
    \end{mdframed}
    \caption{Formal definitions of the problem}
    \label{fig:defs}
\end{figure}

\newcommand{\toolnamectx}{\ensuremath{\text{\toolname}_{\text{Ctx}}}\xspace}
\newcommand{\toolnamesympo}{\ensuremath{\text{\toolname}_{\text{SymPO}}}\xspace}

\section{Method}

\begin{figure*}[t]
    \centering
    \includegraphics[width=.85\linewidth]{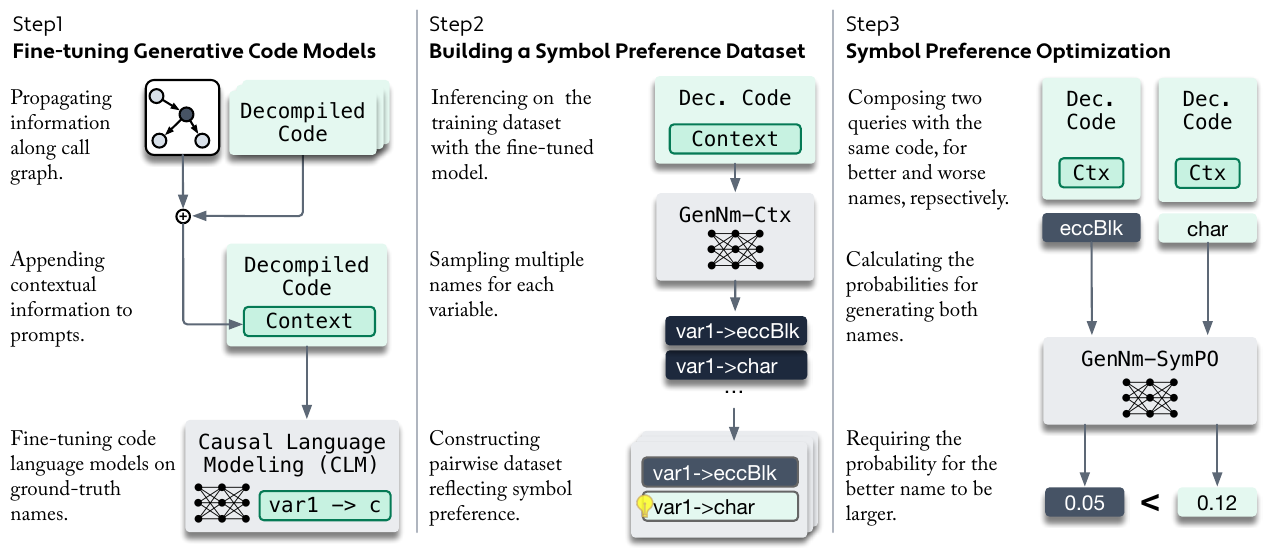}
    \caption{Training pipeline of \toolname}
    \label{fig:workflow}
    \vspace{-5pt}
\end{figure*}

\begin{figure*}[t]
    \centering
    \includegraphics[width=.7\linewidth]{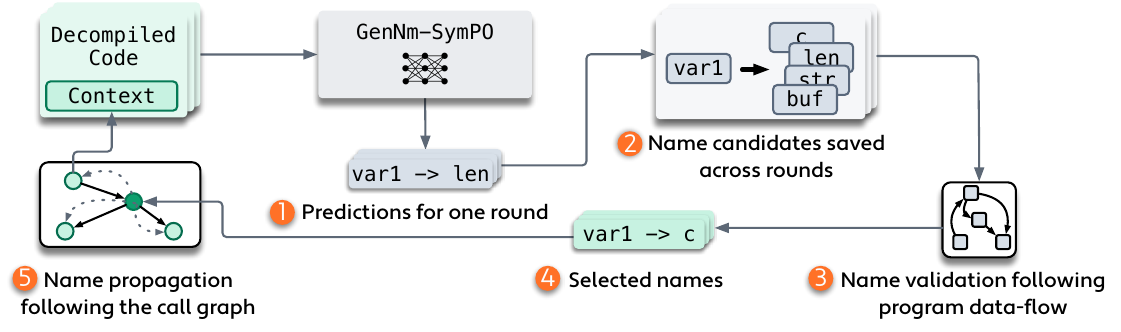}
    \caption{Inference pipeline of \toolname}
    \label{fig:wf-infer}
    \vspace{-5pt}
\end{figure*}

\subsection{Overview}

\noindent
{\bf Training.} We show the training pipeline of \toolname in Fig.~\ref{fig:workflow}. We train \toolname in three steps: (1)~The training process starts from a pre-trained checkpoint of a code language model~(e.g., CodeLlama-7B). It first fine-tunes the pre-trained model on decompiled code to align the distribution of the pre-trained model to the distribution of decompiled code (and the ground-truth names),
resulting in a model noted as \toolnamectx. 
(2)~We use \toolnamectx to inference on the training dataset, and construct a pairwise symbol preference dataset from the model's predictions. Each data sample in the symbol preference dataset contains a preferred name and a less preferred name. (3)~We further train the model with the symbol preference optimization on the preference dataset, resulting in a model noted as \toolnamesympo.

\noindent
{\bf Inference.} The inference process is depicted by Fig.~\ref{fig:wf-infer}. We solve the name recovery problem with an iterative process. At each round, the \toolname model predicts names for individual decompiled functions,
using the global contextual information collected from previous rounds~(Step 1). Then the predictions are added to a candidate name map from a variable to the candidate names of this variable seen across rounds~(Step 2). We 
then leverage the name validation algorithm to select best candidate names based on program data-flow~(Steps~3--4). Finally, the selected names are propagated following the call-graph, updating the quries to the \toolname model~(Step 5) in the next round. It terminates when no variable name is updated or until a predefined budget is reached. 

We discuss the training pipeline in Sections~\ref{sec:method:ft-gen}~and~\ref{sec:method:sympo}. The inference process is discussed in Section~\ref{sec:method:infer}.

\subsection{Fine-tuning Generative Model}
\label{sec:method:ft-gen}

To bridge the gap between the distribution of a pre-trained code language model and the decompiled code,
we fine-tune our model from checkpoints of a pre-trained model~(e.g., CodeLlama-7B). Our fine-tuning  involves two types of datasets: one dataset that contains individual decompiled functions and the corresponding ground-truth variable names and  the other dataset that additionally contains the global contextual information obtained following the program call graph. We fine-tune a model with both datasets because we want our model to have the capabilities of inferring names from local information and
generating names considering global contextual information. The training objective aligns with how the fine-tuned model is used in the inference stage. We leverage the {\em causal language modeling} (CLM)~\cite{radford2018improving} %
loss for fine-tuning. The loss is computed on tokens in both the query decompiled functions and the output names.~\looseness=-1

\noindent
{\bf Dataset w/ local information.}
We note the dataset that contains individual decompiled functions as $D_{\mathrm{loc}}$.
Formally, the dataset $D_{\mathrm{loc}}$ is defined as follows:
\begin{equation}
\label{eqt:data-creat}
\small
\begin{aligned}
    D_{\mathrm{loc}} \Coloneqq \big\{( & \textbf{\textit{query}}:(f.\textit{\textbf{body}}, f.\textit{\textbf{ids}}), \\ & \textbf{\textit{resp}}: n[f.\textit{\textbf{fid}}] )
    \big| (b, n) \in \mathcal{D} \land  f \in b\big\},    
\end{aligned}
\end{equation} where $\mathcal{D}$
denotes the list of binary programs used for training, and $b$ and $n$ a binary and its name map, respectively, as defined in Fig.~\ref{fig:defs}. Hence $n[f.\textit{\textbf{fid}}]$ denotes the map from a placeholder variable name to the ground-truth variable name for function $f$.~\looseness=-1

\noindent
{\bf Context Propagation.}
\revise{
Names in calling contexts can help the model understand the semantics of the function. Intuitively, names from the caller functions may provide hints about the higher-level purpose of the function, and names from the callee functions may provide details about the primitive functionalities of the analyzed function.
}
We first discuss the context propagation algorithm that gathers names following the program call graph, and then discuss how we use it to construct the dataset with additional contextual information. Note that the algorithm is used to construct the contextual dataset during the training time and to propagate and update model query inputs during the inference time. 

The context propagation algorithm takes as input the call graph of a program and the predictions 
for
individual functions and 
propagates the predicted names
along the call graph.
Intuitively, the propagation algorithm gathers information from both the caller functions and the callee functions of an analyzed function $f$. For the caller functions, the algorithm identifies the callsites, i.e., the call expressions that call $f$. It then renames the placeholder names in the corresponding call expressions with the names predicted from the local context of the caller function, and appends the renamed call expressions to the query of $f$. Similarly, the algorithm renames the signature of the callee functions of $f$ and appends them to the query of $f$.

Given a function $f$, we formally define the %
context propagation rules as follows:
\begin{align}
& \begin{aligned}
\small
\label{eqt:prop-caller}
    \textit{CallerCtx}(f, n) \Coloneqq & \bigcup  \big\{
    \textbf{rename}(f.\fieldemph{fid}(a), n[clr.\fieldemph{fid}]) \\ 
    & \big| (clr,f)\in b.\fieldemph{cg} \land f.\fieldemph{fid}(a) \in clr.\fieldemph{body} \big\}
\end{aligned} \\
& \begin{aligned}
\small
\label{eqt:prop-callee}
    \textit{CalleeCtx}(f, n) \Coloneqq & \bigcup \big\{\textbf{rename}(cle.\fieldemph{fid}(r), n[cle.\fieldemph{fid}]) \\
    & \big| (f,cle)\in b.\textbf{\textit{cg}} \land (r,\_)=cle.\fieldemph{body}\big\}
\end{aligned} \\
\small
& \begin{aligned}
\small
\textit{Ctx}(f, n)\! \Coloneqq\! \textit{CallerCtx}(f, n) \cup \textit{CalleeCtx}(f, n) \\
\end{aligned}
\end{align} where $b$ and $n$ are the binary program that the function $f$ belongs to and the corresponding name map. The name map contains the ground-truth names when constructing the training dataset
and the predicted names when propagating names during inference. The utility function $\textbf{rename}(x, y)$ 
renames all $id$s in $x$ according to the name map $y$.

\begin{figure*}[t]
    \centering
    \includegraphics[width=.7\linewidth]{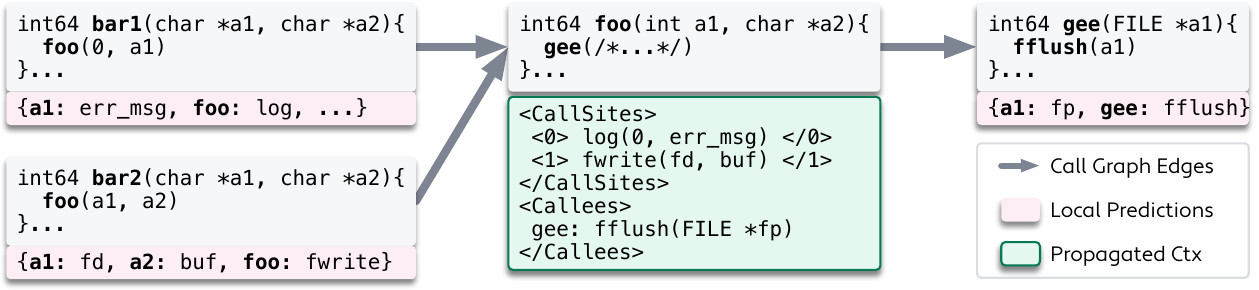}
    \caption{Example of propagating global contextual information along the call graph. \revise{
Initially, \toolname reasons each function independently and obtains the results shown in the pink boxes. After that, \toolname propagates names along the call graph.
The green box under {\tt foo()} shows the propagated contextual information. For example, %
in {\tt bar1()}, the model predicts the names \texttt{err\_msg} and \texttt{log} for {\tt a1} and {\tt foo}, respectively. Therefore, in the context of {\tt foo()} in the middle column, the %
algorithm renames the call statement to {\tt foo()} with the predicted names and propagates it as the $0$-th entry of the callsites.
Similarly, it renames the signature of the callee function {\tt gee()} with the predicted names, and propagates the renamed signature to the analyzed function. 
We can see that the names from caller functions hint the model that the purpose of {\tt foo()} might be writing messages to a file, and the names from the callee function hint the model that {\tt foo()} flushes the output buffer.
}}
    \label{fig:ex-prop}
\end{figure*}

Given a data sample containing a decompiled function $f$,
Equation~\ref{eqt:prop-caller} depicts the rule to propagate names from its caller. Specifically, $(clr, f)\in b.\textbf{\textit{cg}}$ describes the constraint that $clr$ is a caller of $f$, 
and $f.\fieldemph{fid}(a)$ refers to a %
call expression in the body of $clr$ that calls $f$; $f.\fieldemph{fid}$ denotes the placeholder name of $f$ and $a$ denotes the argument list. The propagation algorithm uses names in $n[clr.\fieldemph{fid}]$ to rename the placeholder names in the call expression, and then adds it to the context of $f$.
Similarly, Equation~\ref{eqt:prop-callee} depicts the rule to propagate context from the callee of $f$.
As defined in Fig.~\ref{fig:defs}, $r$ denotes the parameter list of $cle$, and $cle.\fieldemph{fid}$ refers to the placeholder name of $cle$. Therefore,
$cle.\fieldemph{fid}(r)$ denotes the signature of the callee function $cle$. The algorithm renames the placeholder names in the signature of the callee function and adds it to the context of $f$. Fig.~\ref{fig:ex-prop} shows a concrete example.

An alternative design 
is simply appending the bodies of caller and callee functions to a query function. As discussed in Section~\ref{sec:moti-our-tech}, it 
is neither efficient since it significantly increases the number of query tokens nor effective due to the degradation of model's performance with longer input context.

\noindent
{\bf Dataset w/ contextual information.}
We formally define the dataset with contextual information (noted as $D_{\mathrm{ctx}}$) as follows:
\vspace{-3pt}
\begin{equation}
\label{eqt:data-creat}
\small
\begin{aligned}
    D_{\mathrm{ctx}} \Coloneqq \big\{& (\textbf{\textit{query}}:(f.\textit{\textbf{body}}, Ctx(f,n), f.\textit{\textbf{ids}}), \\ & \textbf{\textit{resp}}: n[f.\textit{\textbf{fid}}] )
    \big| (b, n) \in \mathcal{D} \land  f \in b\big\},    
\end{aligned}
\end{equation} where $\mathcal{D}$
denotes binaries used for training, and $b$ and $n$ a binary and its name map, respectively, as defined in Fig.~\ref{fig:defs}; $Ctx(f,n)$ denotes the contextual information gathered by the context propagation algorithm.

\noindent
{\bf Loss function for fine-tuning.} We use a %
CLM loss to fine-tune
on both datasets. The loss is formally defined as follows:
\vspace{-3pt}
\begin{equation}
\small
\label{eqt:clm-loss}
\vspace{-3pt}
\begin{aligned}
    & \mathcal{L}_{\mathrm{ft}}(\Theta, D_{\mathrm{loc}}, D_{\mathrm{ctx}}) \Coloneqq \\
    & 
    -\mathbb{E}_{(q,r)\sim D_{\mathrm{loc}} \cup D_{\mathrm{ctx}}}\Big[
     \sum_{i=1}^{\text{len}(q)+\text{len}(r)}\log P(\mathbf{x}_{i}|\mathbf{x}_{<i};\Theta) \Big],
\end{aligned}
\end{equation}
where $\Theta$ denotes the weights of the fine-tuned model; $\mathbf{x}$ denotes the sequence obtained by concatenating the tokens in the query~($q$) and the tokens in the response~($r$);
$\mathbf{x}_{i}$ denotes the $i$-th token in $\mathbf{x}$; and $\mathbf{x}_{<i}$ the token sequence before the $i$-th token.
Our fine-tuning stage calculates the CLM loss for tokens in both the query and the response to help the model understand the distribution of the decompiled code in the query.~\looseness=-1

\subsection{Symbol Preference Optimization}
\label{sec:method:sympo}

In the natural language domain, \textit{preference} denotes that a natural language sentence output by a 
generative model is preferred over another.
Preference optimization is a method 
to align the behavior of a pre-trained LLM to human preference~\cite{rafailov2024direct}. It takes as input pairwise data samples, and asks a model to predict a higher probability for the preferred response and a lower probability for the less preferred response. 
Since our technique is based on generative models, in order to counter biases, we design a SymPO method for our task.
The SymPO dataset contains pairwise data samples. Each sample consists of a query function, a less preferred name~(indicating the model's biases), and a preferred name. Both 
are sampled from the model's output. Instead of involving a human evaluator, we use the string similarity to the ground-truth name as the preference  for a given variable name.
The SymPO loss is carefully designed so that it teaches the model to select preferred names over the less preferred names while not compromising the model's capability on the variable recovery problem.~\looseness=-1

We first introduce how we construct the pairwise dataset used for SymPO~(i.e., Step 2 in Fig.~\ref{fig:workflow}), and then introduce the SymPO loss~(i.e., Step 3 in Fig.~\ref{fig:workflow}).

\noindent
{\bf Constructing the SymPO dataset.} We construct the dataset using  \toolnamectx to inference a subset of the training data, and sampling the top 20 predictions from the model for each query. We collect cases where \toolnamectx makes mistakes but has at least another response in the top 20 predictions that is significantly better. Intuitively, the model has the knowledge of 
better names for those cases, yet it makes mistakes due to the biases. The SymPO process thus has the chance to fix the biases without changing the model significantly.

An alternative design is to use the ground-truth as the preferred names. However, the results in Section~\ref{sec:eval:ablation} show that using ground-truth names underperforms compared to
using the best predictions of \toolnamectx as the preferred names. 
That is because \toolnamectx may not 
learn how to generate the ground-truth names 
for certain programs.
Cases where the ground-truth diverge too much from \toolnamectx's learned distribution negatively affect the model's performance.

We formally present 
the SymPO dataset as follows.
First, we use $\Hat{D}$ to denote the inferenced training subset.
\vspace{-3pt}
\begin{equation}
\label{eqt:data-pred}
\small
\begin{aligned}
    \Hat{D} \Coloneqq \big\{(\fieldemph{query}: q,\ &\fieldemph{preds}: \hat{r},\ \fieldemph{gt}: r )\big| (q, r) \in \mathcal{D}_{\mathrm{loc}} \cup \mathcal{D}_{\mathrm{ctx}} \\
    & \land  \hat{r} = \text{\toolname}_{\mathrm{ctx}}(q, \text{top20}) \big\},    
\end{aligned}
\end{equation}
where $\text{\toolname}_{\mathrm{ctx}}(q, \text{top20})$ denotes the top 20 responses returned by \toolnamectx given a query $q$.

The SymPO dataset, noted as $D_{\mathrm{prf}}$, is defined as follows:
\vspace{-3pt}
\begin{equation}
\label{eqt:data-pref}
\small
\begin{aligned}
    D_{\mathrm{prf}} \Coloneqq \big\{ & (\fieldemph{query}:q,\  
    \fieldemph{better}: r_b,\ \fieldemph{worse}: r_w) \\
    & 
    \big| (q, \hat{r}, r) \in \Hat{\mathcal{D}} \land  r_w = \textbf{sample}(\hat{r}) \\
    & \land r_b = \textbf{best}(\hat{r}, r) \land r_b \succ_r r_w \big\},    
\end{aligned}
\end{equation}
where $\textbf{best}(\hat{r}, r)$ denotes the name in $\hat{r}$ that is most similar to the given ground-truth name $r$,
$\textbf{sample}(\hat{r})$ denotes a name that is randomly sampled from $\hat{r}$,
and $r_b \succ_r r_w$ denotes that the name $r_b$ is significantly more similar to the given ground-truth name $r$ than $r_w$. \revise{We use token-level precision and recall to measure the similarity between a predicted name and the ground-truth name.}

Moreover, to reduce the noise in the SymPO dataset and improve the training efficiency, we use lightweight static code features as heuristics to filter out low-quality data. Empirically, our static heuristics reduce the dataset size by 60\%, and results in Section~\ref{sec:eval:ablation} show that the performance achieved by training on the reduced dataset is even slightly better than training on all the data samples. 
\revise{
In particular, we remove a function if more than two-thirds of its callee functions do not have meaningful names. Optimizing model's preference on those data samples introduces only noises because the local information may not be enough for the model to predict good names. In addition, we remove functions with less than 5 statements and meanwhile do not have branches. Note that we only remove functions for constructing the SymPO training dataset. We do not remove any functions from the test dataset.
}

\noindent
{\bf Loss function of SymPO.} 
The loss function of SymPO is adapted from the loss function proposed in direct preference optimization~\cite{rafailov2024direct}, which is used to align human preference with fine-tuned LLMs. The loss function has two sub-goals: (1)~guiding the model to generate better names with higher probabilities (than the probabilities for generating worse names), and (2)~preventing the model from diverging too much from its original distribution.
The loss function is formally presented as follows:
\begin{equation}
\label{eqt:loss-sympo}
\footnotesize
\begin{aligned}
& \mathcal{L}_{\mathrm{SymPO}}(\Theta, \Theta_{\mathrm{ctx}}) \Coloneqq \\
& -\!\mathbb{E}_{(q,b,w)\sim D_{\mathrm{prf}}}
\!\Bigg[\!
\log\sigma\Big(
\beta\! \log \frac{\mathbf{P}(b|q;\Theta)}{\mathbf{P}(b|q;\Theta_{\mathrm{ctx}})}
\!-\!
\beta\! \log \frac{\mathbf{P}(w|q;\Theta)}{\mathbf{P}(w|q;\Theta_{\mathrm{ctx}})}
\Big)\Bigg]
\end{aligned}
\end{equation}
where $\Theta$ and $\Theta_{\mathrm{Ctx}}$ denote the weights of \toolnamesympo and the weights of \toolnamectx, respectively. $\beta$ is a hyper-parameter that controls the sensitivity of the loss to the margin between the probability for better names and the probability for worse names. The loss is optimized w.r.t. $\Theta$ only. In other words, the weights of \toolnamectx are frozen during SymPO.

\begin{figure}
    \centering
    \includegraphics[width=.7\linewidth]{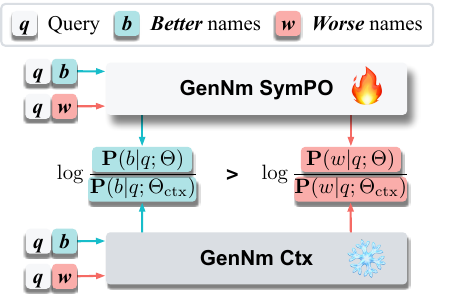}
    \caption{Loss for SymPO. The weights of \toolnamectx are frozen. The weights of  \toolnamesympo are optimized guided by the SymPO loss. Preferred (better) names and the corresponding probabilities are in blue; less preferred (worse) ones are in red.}
    \label{fig:loss-sympo}
\end{figure}

An intuitive explanation for the loss function is visualized in Fig.~\ref{fig:loss-sympo}. Two models are involved in SymPO. The first model is \toolnamesympo, which is optimized by the loss function. It is initialized with the weights of \toolnamectx. The other model is a frozen \toolnamectx, which will not be updated during training. It is used as a ``reference'' model so that the divergence of \toolnamesympo is constrained. A detailed discussion for the loss function is in Appendix~\ref{appdx:sympo-gradient-analysis}. Assume a data sample consisting of the query function~($q$), a better name~($b$), and a worse name~($w$). The blue parts in Fig.~\ref{fig:loss-sympo} depict the first loss term in Equation~\ref{eqt:loss-sympo} (i.e., $\log \frac{\mathbf{P}(b|q;\Theta)}{\mathbf{P}(b|q;\Theta_{\mathrm{ctx}})}$). It uses both models to calculate the probabilities of generating the better name $b$ and guides  \toolnamesympo to produce a larger probability for $b$ than  \toolnamectx. Similarly, the red parts~(corresponding to the second loss term) require \toolnamesympo to generate a significantly smaller probability compared to the \toolnamectx.

\subsection{\revise{Context Augmentation at the Inference Stage}}
\label{sec:method:infer}

\revise{
At the inference stage, we iteratively run \toolname because the input contexts provided to the models are updated based on the latest round of predictions. In each iteration, the newly generated names along with the names generated in the previous rounds are considered candidate names for the variable. We propagate names along the program call graph to provide contextual information. The algorithm ~(Step 5 in Fig.~\ref{fig:wf-infer}) is discussed in previous sections.

To select the final name prediction across different iterations, \toolname leverages program analysis to aggregate the names predicted in different iterations, and selects the candidate name with the maximum level of consistency. The analysis in \toolname is a general data-flow analysis customized to the domain of variable names, and the consistency check is achieved by majority voting. The implementation details of both techniques are in Appendix~\ref{appdx:pa-and-voting}. Note that there is existing work~\cite{resym} exploring the combination of program analysis and LLM at the inference stage. Therefore, although our analysis is customized to the variable name domain and is different with existing work~\cite{resym}, we do not claim conceptual novelty for the analysis and voting algorithm.
}

%% file: evaluation.tex
\section{Experimental Setup}
\label{sec:setup}
\subsection{Dataset}

We evaluate \toolname on two commonly used~\cite{varbert,dire,dirty} datasets. The first one is built following the same process as DIRTY~\cite{dirty} (noted as the DIRTY dataset) and the other is derived from the released VarCorpus dataset used by VarBERT~\cite{varbert} (noted as the VarCorpus dataset). The DIRTY dataset is built from popular GitHub projects, and the VarCorpus dataset is built~(by the VarBERT authors) from a Linux package manager, Gentoo~\cite{gentoo}. We rebuild the DIRTY dataset  because the original DIRTY dataset contains binary programs that are not fully stripped~\cite{varbert}. Additionally, the dataset provided by DIRTY's authors contains only preprocessed data without raw binaries. Our technique requires call graphs of programs and thus cannot directly use the provided DIRTY dataset. For the VarCorpus dataset, thanks to the help of VarBERT's authors, we obtain the corresponding binary programs in VarCorpus and thus can reuse the processed VarCorpus dataset with the call graphs extracted from the binary programs. For both datasets, the ground-truth variable names are obtained from the debug information in binary programs. 
For the DIRTY dataset, we reuse the code provided by DIRTY's authors to collect the ground-truth. For the VarCorpus dataset, we directly reuse the ground-truth provided in the dataset.~\looseness=-1

\noindent
{\bf Data quality.} To prevent the data duplication problem as observed by the previous work~\cite{varbert}, we ensure the high quality of both datasets with strict deduplication rules, only including a binary program if at least 70\% of its functions are not seen. In the deduplication process, we conservatively consider two functions as the same functions if they have the same name. We discuss the rationale in Appendix~\ref{appdx:dataset}. As a result, our processed datasets are more diversified than the existing datasets. For example, only 46\% of functions in the original VarCorpus dataset have unique names, indicating that the other 54\% of functions may have similar semantics (an example is in Fig.~\ref{fig:appdx:dedpu-str} in the Appendix). On the other hand, 81.3\% and 89.4\% of functions in our processed VarCorpus and DIRTY datasets have unique names, respectively. Our processed DIRTY and VarCorpus datasets have 348k and 895k functions, respectively. Please see Appendix~\ref{appdx:dataset} for detailed statistics.

\noindent
{\bf Preventing data leakage.} Moreover, we use string similarity-based rules to filter out the overlap between training and test data, preventing potential data leakage. Previous works~\cite{varbert,dirty} use exact string match as the criterion for checking data leakage. However, as shown in Fig.~\ref{fig:appdx:dedpu-str} and Table~\ref{tab:appdx-overlap-perf} in the Appendix, there might be potential data leakage even if the strings of two functions are not exactly the same (e.g., two functions may differ in only one number). To better measure the generalizability of models, we conservatively filter out those potential leakage by filtering out a test sample if its string similarity score to a training sample ~(from 0 to 100) is higher than 90.

\revise{
\noindent
{\bf Data availability.} We submitted our artifact to the artifact evaluation track. We will publish our data splits, model checkpoints, and implementations upon publication. 
}

\subsection{Splits}

For most experiments in the evaluation, we split both datasets with a ratio of 9:1 by binaries (not by functions) for training and test. We randomly sample 5\% functions from the training datasets as the validation sets. We split our training and test datasets by binary programs (instead of by binary functions). That is because splitting data by functions may cause data leakage. Decompilers typically use the address of a global variable or a function to construct a placeholder name for it.  For example, assume two functions from a binary program, and both of them use a global variable {\tt qword\_409abc}. One of the functions is in the training dataset, and hence the training process exposes the ground-truth name, e.g., {\tt message}, to the model. During test, the model can easily predict {\tt qword\_409abc} as {\tt message} since the placeholder name is already seen in the training data. 
\revise{To fairly compare the improvements achieved by \toolname with the baseline techniques, we additionally conduct experiments with the split-by-function setup following the previous work~\cite{varbert} in Section~\ref{sec:sem-match} . 
}

\subsection{Models}

\revise{
Due to limited resources, we train \toolname from CodeGemma-2B~\cite{team2024codegemma} for most of the experiments. To study how different sizes of models may affect the performance, we additionally train two \toolname models from CodeLlama-7B and CodeLlama-34B~\cite{codellama} on the DIRTY dataset.
The detailed hyperparameters of our model are listed in Table~\ref{tab:eval:hyper-params} of the Appendix.
We use VarBERT~\cite{varbert} and ReSym~\cite{resym} as the baseline techniques. VarBERT~\cite{varbert} is a representative classification based method that demonstrates better performance than previous state-of-the-art models~\cite{dire,dirty}.
ReSym~\cite{resym} is a recent technique based on LLM.
It also demonstrates better performance than previous state-of-the-art models~\cite{dire,dirty}.
We train all models until they converge (i.e., the validation loss no longer decreases). We select models that achieve the best validation loss.
}

\subsection{Metrics}

We use two sets of metrics to evaluate model performance.

\noindent
{\bf Token-based semantics match.} Previous works use exact string match to evaluate the performance of a variable name recovery technique. However, exact string match cannot faithfully reflect the capability of a tool. As discussed in SymLM~\cite{symlm}, a previous work focusing on recovering function names, even when two variables have the same meaning, the names specified by developers may vary due to many reasons, e.g.,  use of abbreviations and concatenation of names.
We thus adapt the same metrics used in SymLM to measure the quality of generated names.
Intuitively, given a ground-truth name $n$ and a predicted name $\hat{n}$, the metric tokenizes both names into sets of tokens, noted as $W$ and $\Hat{W}$. Then it uses set comparison to calculate precision and recall. Formally,
\newcommand{\twobars}{\ensuremath\big|\!\big|\xspace}
\begin{equation}
\footnotesize
\label{eqt:eval-pr}
    Precision(W, \Hat{W}) = \frac{\twobars\{\hat{w}|\hat{w} \in \Hat{W} \land \exists w \in W, \hat{w} \simeq w\}\twobars}{\twobars \Hat{W} \twobars}
\end{equation}
\begin{equation}
\footnotesize
\label{eqt:eval-rc}
    Recall(W, \Hat{W}) = \frac{\twobars\{\hat{w}|\hat{w} \in \Hat{W} \land \exists w \in W, \hat{w} \simeq w\}\twobars}{\twobars W \twobars}
\end{equation}
In Equations~\ref{eqt:eval-pr}~and~\ref{eqt:eval-rc}, $\simeq$ denotes whether two tokens have similar semantics. SymLM~\cite{symlm} built a semantics word cluster trained on CodeSearchNet~\cite{husain2019codesearchnet} and derived edit-distance-based rules to measure the semantic similarity between tokens. We reuse their word cluster and rules.

\noindent
{\bf GPT4Evaluator.} Token-based metrics may not accurately reflect whether a name matches the program context or developers' intention. For example, the names {\tt wait\_sec} and {\tt timeout} have no token overlap but denote similar semantics. On the other hand, existing work~\cite{su2024source} on decompiled code summarization demonstrates that using GPT4 as an evaluator aligns better with human judgments than automatic metrics. Therefore, we adapt their method, further using GPT4 as an evaluator to measure the quality of generated names.

Specifically, we follow~\cite{su2024source} and measure the quality of a name from {\it context relevance} and {\it semantics accuracy}. We query GPT4 per binary function. Each query consists of a decompiled function with the ground-truth variable names, and a name map 
from ground-truth names to predicted names. We ask GPT4 to first summarize the decompiled function, and evaluate each predicted name by answering two questions in scores from 1~(worst) to 5~(best): (1) Whether the predicted name is consistent with the program context? (2) Whether the predicted name accurately depicts the semantics of the variable? The prompt used and examples for each score are shown in Fig.~\ref{fig:prompt-gpt4eval} and Fig.~\ref{fig:example-preds} in the Appendix.

\section{Evaluation}

\subsection{Performance in terms of Semantics Match}
\label{sec:sem-match}

\begin{table*}[t]
    \caption{\revise{Performance of \toolname compared with VarBERT and ReSym. Proj. NIT~(Project Not-In-Train) denotes test programs whose corresponding {\it projects} are not seen in the training dataset. Proj. IT~(Project In-Train) denotes test programs whose  projects are seen in the training dataset. \textbf{\em Both Proj. NIT and Proj. IT samples do not overlap with training data samples.}~\looseness=-1 }}
    \label{tab:eval-performance}
    \centering
    \begin{tabular}{cc cc c cc c cc }
    \toprule
    \multirow{2.5}{*}{Dataset} & \multirow{2.5}{*}{Model} & \multicolumn{2}{c}{Proj. NIT} & & \multicolumn{2}{c}{Proj. IT} && \multicolumn{2}{c}{Overall} \\
    \cmidrule{3-4} \cmidrule{6-7} \cmidrule{9-10}
            &       &   Precision & Recall &              & Precision & Recall  && Precision & Recall \\
    \midrule
    \multirow{3}{*}{DIRTY}      & VarBERT      & 23.6 & 21.7 && 31.4 & 29.6 && 27.2 & 25.5 \\ 
& \revise{ReSym}             & \revise{25.3} & \revise{24.9} && \revise{35.6} & \revise{34.3} &&  \revise{30.2} & \revise{29.3} \\
& \toolname & \textbf{30.5} & \textbf{28.8} && \textbf{41.7} & \textbf{39.6} && \textbf{35.8} & \textbf{33.9} \\
    \midrule
    \multirow{3}{*}{VarCorpus} & VarBERT      & 20.9 & 19.3 && 32.5 & 31.0 && 29.8 & 28.3 \\
& \revise{ReSym} & \revise{23.5} & \revise{24.1} && \revise{34.2} & \revise{35.8} && \revise{31.7} & \revise{33.1} \\
& \toolname & \textbf{29.5} & \textbf{27.4} && \textbf{44.7} & \textbf{42.8} && \textbf{41.2} & \textbf{39.3} \\
    \midrule
    \revise{\multirow{3}{*}{
    \begin{minipage}{.1\linewidth}
        \centering
        VarCorpus\\
        Split by Function
    \end{minipage}
    }}
    & \revise{VarBERT} & -  & -  && - & - && \revise{50.0} & \revise{49.2} \\
    & \revise{ReSym} & -  & -  && - & - && \revise{51.2} & \revise{52.2} \\
    & \revise{\toolname} & -  & -  && - & - && \revise{\textbf{62.4}} & \revise{\textbf{62.8}} \\
    \bottomrule
    \end{tabular}
\end{table*}

\revise{
\noindent
{\bf Overall.}
We show the performance of \toolname compared with the baseline techniques in Table~\ref{tab:eval-performance}. We can see that overall, \toolname outperforms both VarBERT and ReSym on all datasets/splits. On the DIRTY dataset, \toolname outperforms VarBERT by 8.5 percentage points in terms of both precision and recall; it outperforms ReSym by 5.6 and 4.6 percentage points in terms of precision and recall, respectively.
On the VarCorpus dataset, \toolname outperforms VarBERT by 11.4 and 11.0 percentage points in terms of precision and recall, respectively; it outperforms ReSym by 9.5 and 6.2 percentage points in terms of precision and recall, respectively. }Note that the performance for VarBERT reproduced in Table~\ref{tab:eval-performance} is lower than the reported statistics in the VarBERT paper. That is expected because we preclude potential overlaps between training and test sets with a stricter setup. Appendix~\ref{appdx:dataset} shows that both \toolname and VarBERT achieve significantly higher performance on the subset of samples that have a high similarity to the training dataset ~(e.g., VarBERT and \toolname achieve a precision of 50.8\% and 72.3\% on the DIRTY dataset, respectively).~\looseness=-1

\noindent
{\bf Project-in-train/project-not-in-train.} Moreover, we observe that complex projects typically contain more than one binary. Different binaries in a project likely share similar coding styles or naming preferences. Therefore, a model may be able to predict better names if the corresponding project of a test program has been seen in the training dataset. Therefore, we further categorize the test programs by whether the corresponding projects are seen during training or not, noted as {\it project-in-train} and {\it project-not-in-train}. Note that this categorization is \textbf{\textit{different}} from the {\it in-train} and {\it not-in-train} setup in DIRTY~\cite{dirty}. As pointed out by previous work~\cite{varbert}, there are better solutions for renaming variables in functions that overlap with the training dataset~(i.e., the ``in-train'' samples in DIRTY's setup). On the other hand, in our setup, {\it project-in-train} mimics a realistic scenario that the naming style of an author group~(e.g., an APT group~\cite{aptgroup}) is already learned beforehand, and a technique is used to analyze programs from the same author group. Both project-in-train and project-not-in-train samples  \textbf{\em do not overlap with the training data samples.}

\revise{
We can see that all techniques perform better on samples whose projects have been seen during training. On those samples, \toolname outperforms VarBERT by more than 10 percentage points (on both datasets) in terms of both precision and recall, and it outperforms ReSym by more than 5 and 7 percentage points on the DIRTY dataset and the VarCorpus dataset, respectively. For the more challenging project-not-in-train samples, \toolname consistently outperforms both baseline techniques by 3.9--8.6 percentage points, demonstrating better generalizability.

\noindent
{\bf Split by function.}
Moreover, following previous work~\cite{varbert}, we further run all techniques on the VarCorpus dataset with the split-by-function setup.
Split-by-function denotes the setup where some functions in a binary are in the training dataset while other functions are in the test dataset.
We randomly sample 15\% binaries from VarCorpus due to limited resources, following the practice of previous work~\cite{varbert,dirty}. All techniques perform significantly better with the split-by-function setup. Especially, we can see that \toolname outperforms both baseline techniques by more than 10 percentage points. That is because the training paradigm and inference stage of \toolname enables it to leverage contextual information. In the split-by-function setup, the caller and callee functions of an analyzed function may already be seen during training. They provide higher quality contextual information than the caller/callee functions in the split-by-binary setup. Therefore, the performance of \toolname improves significantly. It demonstrates the effectiveness of leveraging calling contexts in the name recovery problem.

\noindent
{\bf Significance of improvements.}
Note that the scale of improvement introduced by \toolname over the baselines is comparable to that in existing work.
In the most challenging setup (split by binary, without overlap with training dataset), \toolname outperforms the baseline techniques by 4.6--11.4 percentage points. DIRTY~\cite{dirty} improves over its baseline by 5.1 percentage points (on the DIRTY dataset), and VarBERT~\cite{varbert} improves over its baseline by 4.5 percentage points (on the VarCorpus dataset).~\looseness=-1
In the split-by-function setup, \toolname improves over the baseline techniques by 10.6--13.6 percentage points. VarBERT~\cite{varbert} improves over its baseline by 12.7 and 14.8 percentage points. 

}

\subsection{\revise{Performance w.r.t. Different Sizes of Base Models}}
\label{sec:eval-size}

\revise{

\begin{table}[t]
    \caption{\revise{Performance w.r.t. different sizes of base models.} }
    \label{tab:eval-performance-size}
    \centering
    \begin{tabular}{c cc c cc c cc }
    \toprule
    \multirow{2.5}{*}{Base Model} & \multicolumn{2}{c}{Proj. NIT} & & \multicolumn{2}{c}{Proj. IT} && \multicolumn{2}{c}{Overall} \\
    \cmidrule{2-3} \cmidrule{5-6} \cmidrule{8-9}
                   &   PR & RC &              & PR & RC  && PR & RC \\
    \midrule
    CodeGemma-2B & 29.7 & 28.0 && 38.5 & 36.7 && 33.7 & 32.0 \\
    CodeLlama-7B & 29.9 & 28.8 && 36.7 & 35.5 && 33.1 & 31.9 \\
    \revise{CodeLlama-34B*} & \revise{\textbf{35.9}} & \revise{\textbf{33.4}} && \revise{\textbf{39.5}} & \revise{\textbf{37.4}} && \revise{\textbf{37.1}} & \revise{\textbf{35.3}} \\
    \midrule
    \multicolumn{9}{l}{\scriptsize {*We fine-tune CodeLlama-34B with LoRA.}} \\
    \bottomrule
    \end{tabular}
\end{table}}

\revise{
\toolname fine-tunes pre-trained code language models. To study how base models with different sizes affect the performance, we additionally train \toolname with CodeLlama-7B and CodeLlama-34B~\cite{codellama}. Note that our resource cannot support a fully fine-tuning for the 34B model. Therefore, we use LoRA~\cite{hu2021lora} to fine-tune the 34B model.
We evaluate all models on a subset of the DIRTY dataset. The results are shown in Table~\ref{tab:eval-performance-size}. We can see that \toolname fine-tuned from CodeLlama-34B achieves significantly better results than \toolname on CodeGemma-2B and CodeLlama-7B. Especially, for the most challenging setup where the project of a binary is not seen in the training dataset, the 34B version of \toolname outperforms the other versions by around 5 percentage points in both precision and recall. That demonstrates the training paradigm of \toolname can generalize to larger models.
}

\subsection{\revise{Generalization to Different Compiler Optimizations}}
\label{sec:eval-opt}

\begin{figure*}[t]
    \centering
    \includegraphics[width=.8\linewidth]{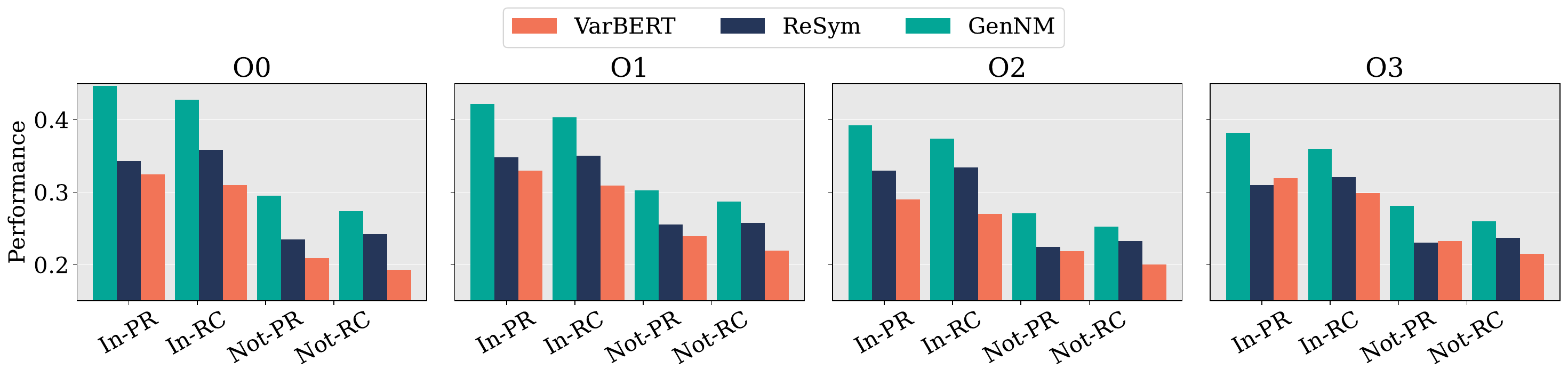}
    \caption{\revise{Generalizability to other optimization levels. \textit{In-PR}, \textit{In-RC}, \textit{Not-PR}, and \textit{Not-RC} denote the average \textit{precision} and \textit{recall} on samples whose project is seen or not seen in the train data, respectively.}}
    \label{fig:eval-gen-to-other}
\end{figure*}

\revise{
To evaluate the generalization of \toolname to other compiler optimization levels, we compare \toolname with both baseline techniques on programs compiled with different  optimization levels from {\tt -O0} to {\tt -O3}. The results are shown in Fig.~\ref{fig:eval-gen-to-other}. We can see that \toolname outperforms both baselines across all optimization levels. It demonstrates that \toolname can generalize to optimized programs. The improvements of \toolname on programs compiled with less aggressive optimizations (i.e., {\tt -O0} and {\tt -O1}) are more significant than the improvements on programs compiled with {\tt -O2} and {\tt -O3}. That is because programs compiled with aggressive optimizations are significantly longer and diverge further from the distribution of source code. Therefore, it is more challenging for models to understand them, affecting the model's performance.
We leave it as future work to further improve the model's capability of understanding programs compiled with aggressive optimization flags.
}

\subsection{\revise{Generalization to Rare Names}}
\label{sec:perf-general}

\begin{figure}[t]
    \centering
    \includegraphics[width=.75\linewidth]{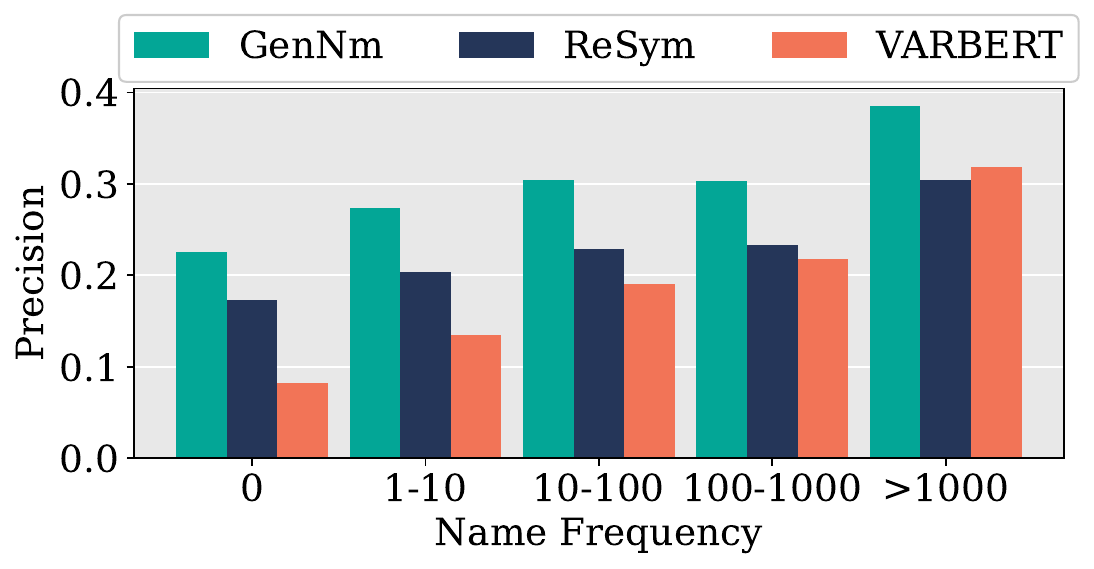}
    \caption{Performance by name frequency on VarCorpus. The x-axis denotes the frequency of the ground-truth name for a variable in the training dataset of VarCorpus, and the y-axis the average precision achieved on the corresponding variables.~\looseness=-1}
    \label{fig:perf-by-freq}
\end{figure}

\revise{
We show \toolname generalizes better to rare names in Fig.~\ref{fig:perf-by-freq}.
Observe that all techniques achieve better performance on names that appear more frequently in the training dataset, and \toolname consistently outperforms both baseline techniques on names with all name frequencies. Moreover, \toolname is more robust when the frequencies of names decrease. For names that are never seen in the training dataset, both \toolname and ReSym outperform VarBERT. Especially, \toolname achieves a precision of over 20\%, which is close to 2 times the performance of VarBERT on those variables. It supports that the generative model generalizes better than a classification model on unseen names.

The performance of \toolname on rare variables~(i.e., variables with a  name frequency from 1 to 10) is 27.1\%, while the performance of VarBERT and ReSym are 13.5\% and 20.4\%, respectively.
That indicates \toolname mitigates the biases of frequent names in the training dataset.
Moreover, we show that 95\% of the rare names are composed of frequently appeared tokens. Details are in Fig.~\ref{fig:tok-freq} in the Appendix.
}

\subsection{Performance Evaluated by GPT4Evaluator}

\begin{figure}[t]
    \centering
    \includegraphics[width=.85\linewidth]{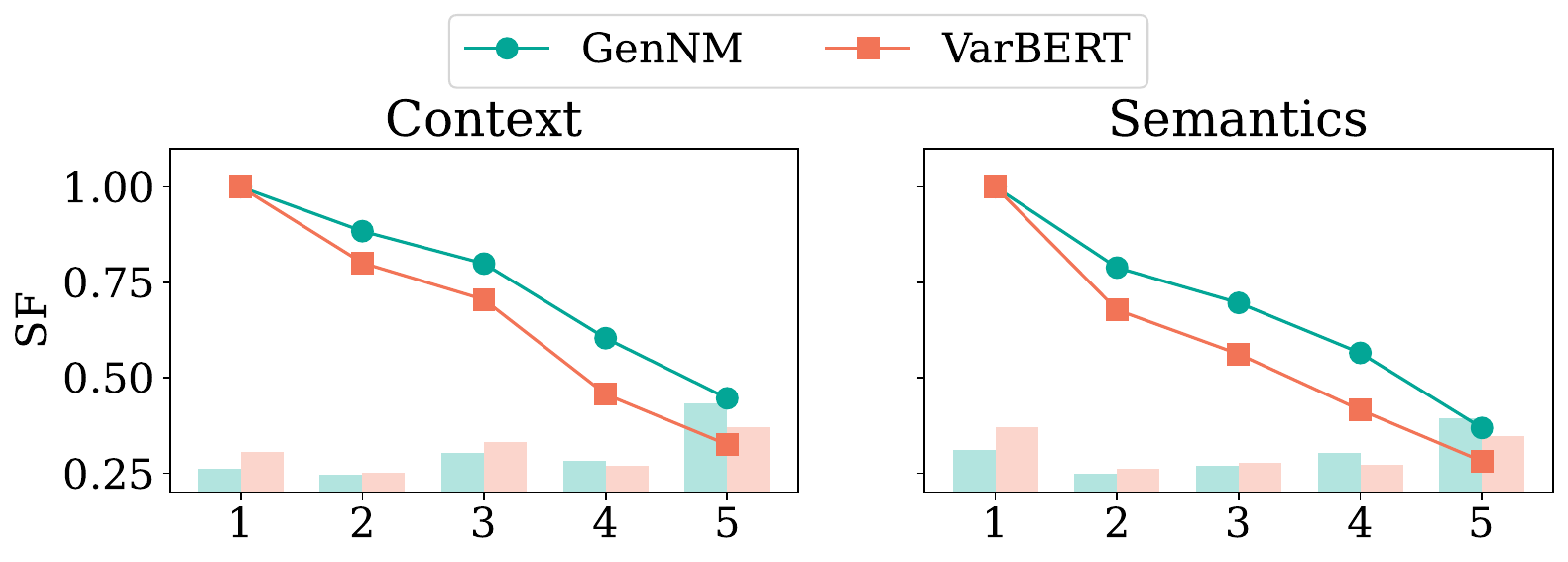}
    \caption{Performance evaluated by the GPT4Evaluator. The two sub-figures show the scores for context relevance (Context) and semantics accuracy (Semantics), respectively.
    SF denotes the survival function. It indicates the number of samples achieving at least the corresponding score. The transparent bars reflect the distribution for each score. }
    \label{fig:perf-gpt4eval}
\end{figure}

We further use GPT4Evaluator to evaluate the performance of both models. Due to the limited budget, we randomly sample 500 functions (corresponding to 1632 variable names) from the DIRTY dataset. The results are shown in Fig.~\ref{fig:perf-gpt4eval}. We can see that in terms of both \textit{context relevance} and \textit{semantics accuracy}, \toolname achieves better scores than VarBERT. Especially, observe that for more than 50\% of variables, the names generated by \toolname are given scores of 4 or better for both measurements, indicating that \toolname can effectively recover high-level semantics information from decompiled code. Fig.~\ref{fig:example-preds} in the Appendix shows examples for names with different scores.
It is also worth noting that \toolname performs better in terms of context relevance than semantics accuracy. It indicates that \toolname can predict names within the correct program context most of the time, 
yet it is more challenging to generate names that accurately reflect the semantics of ground-truth names. 
That is because compared to predicting names that are consistent with the program context, predicting the precise semantics of a variable entails a more accurate understanding of the semantics of the program, which is a challenging problem when the program does not have meaningful symbols~\cite{su2024codeart}. We leave as future work to further improve the model's understanding of decompiled code.

\subsection{Performance Compared to Blackbox LLMs}
\label{sec:eval:compare-llm}

\begin{table}[t]
    \caption{Performance compared to blackbox LLMs.}
    \label{tab:eval-blackbox-llm}
    \centering
    \begin{tabular}{cc cc}
    \toprule
    Model & Prompt & Precision & Recall \\
    \midrule
    \multirow{2}{*}{GPT-3.5} & zero-shot & 26.2 & 27.7 \\
                             & 3-shot & 29.7 & 28.9 \\
    \multirow{2}{*}{GPT-4}  & zero-shot & 30.3 & 33.3 \\
                             & 3-shot & 31.4 & 32.6\\
     \multirow{2}{*}{\revise{CodeLlama-70B}}  
     & \revise{zero-shot} & - & - \\
     & \revise{3-shot} & \revise{27.4} & \revise{26.9} \\
     \midrule
     \toolname & - & \textbf{42.7} & \textbf{39.7} \\
    \bottomrule
    \end{tabular}
\end{table}

We compare the performance of \toolname with LLMs used as black-boxes. We randomly sample 1000 functions from the DIRTY dataset and query two state-of-the-art black-box LLMs (i.e., GPT-3.5 and GPT-4) and one large code LLM (i.e., CodeLlama-70B), with both the zero-shot and 3-shot setups. The prompts used are shown in Appendix~\ref{appdx:prompt-gpt}.
The results are shown in Table~\ref{tab:eval-blackbox-llm}. Observe that GPT-4 achieves better performance than GPT-3.5, and both LLMs achieve better performance in the 3-shot setup. However, due to the distribution gap between decompiled functions and the pre-training knowledge of LLMs, both models underperform \toolname. \toolname outperforms the best results achieved by black-box LLMs by 11.3 and 6.4 percentage points in terms of precision and recall, respectively. 
\revise{For the code LLM, we observe most of its outputs have format errors in the zero-shot setup. We thus only calculate its performance on the 3-shot experiment. Note that it achieves results close to GPT-3.5 but inferior to GPT-4. We speculate that it is because the training data of GPT-3.5 and GPT-4 also contain significant amount of code. Therefore, the LLM specially trained on code may not have advantage given that its size is much smaller than GPT-4.}
It is worth noting that all LLMs are likely to be significantly larger in size~(i.e., 10x--100x)  than the base model used in \toolname. It demonstrates the necessity and effectiveness of fine-tuning a pre-trained code language model on this task.

\begin{figure}
    \centering
    \includegraphics[width=0.5\linewidth]{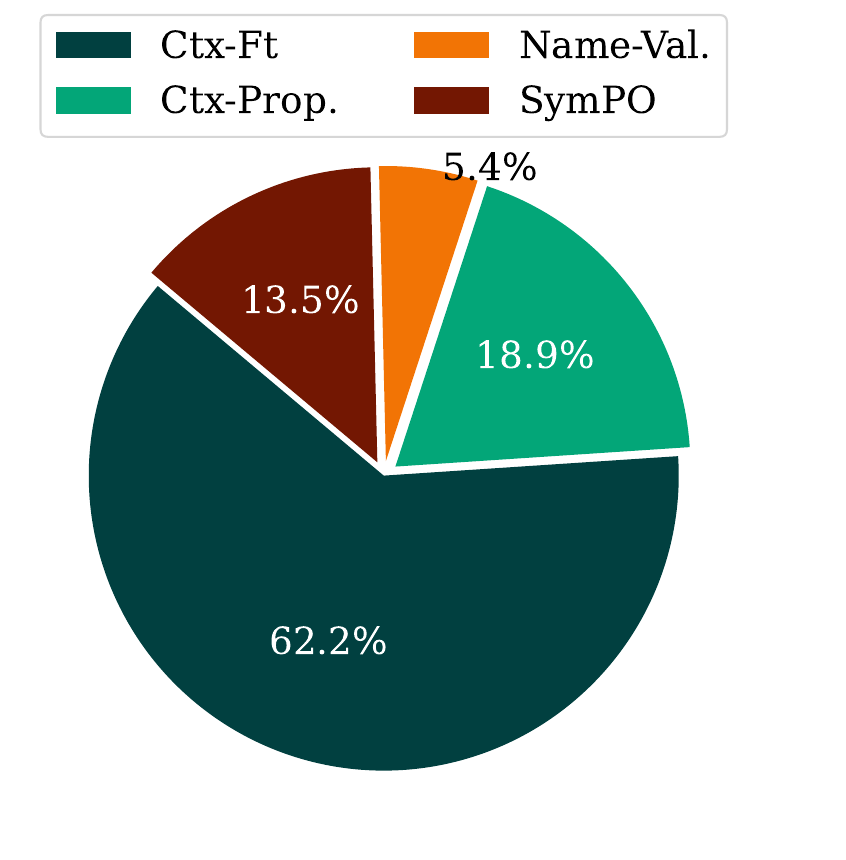}
    \caption{\revise{Attribution of the improvements over ReSym to different components. {\it Ctx-Ft} and {\it SymPO} denote using the context-aware fine-tuning paradigm and the SymPO objective at the training stage, respectively. 
    {\it Ctx-Prop.} and {\it Name-Val.} denote using the context propagation algorithm and the name validation algorithm at the inference stage, respectively.}}
    \label{fig:eval-contrib}
\end{figure}

\begin{figure*}[t]
    \centering
    \begin{subfigure}[t]{.43\linewidth}
    \centering
        \includegraphics[width=\linewidth]{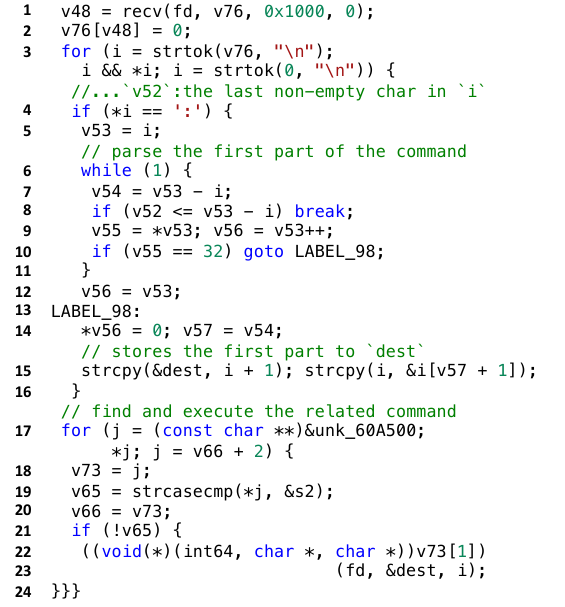}
        \caption{Decompiled code output by IDA}
        \label{fig:mal-decomp}
    \end{subfigure}
    \hspace{0.03\linewidth}
    \begin{subfigure}[t]{.43\linewidth}
    \centering
        \includegraphics[width=\linewidth]{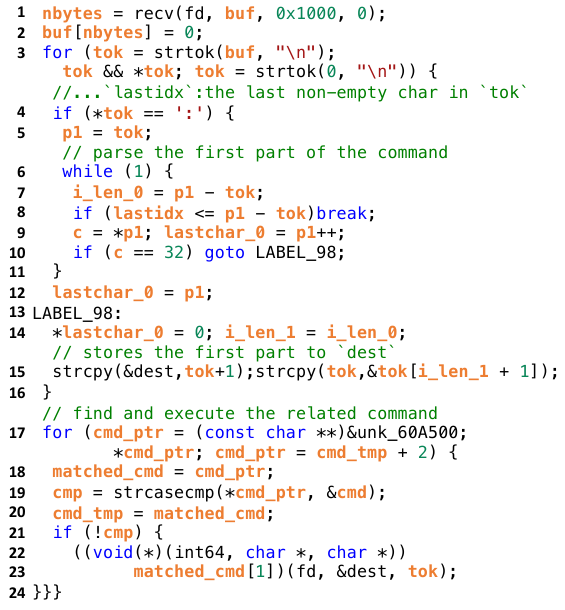}
        \caption{Renamed code (generated names highlighted in orange)}
        \label{fig:mal-rename}
    \end{subfigure}
    \caption{How \toolname helps security analyst understand a malware sample}
    \label{fig:malware-app}
\end{figure*}

\subsection{Ablation Study}
\label{sec:eval:ablation}

We conduct ablation studies to analyze how each component contributes to the effectiveness of \toolname. Moreover, we study the effectiveness of different design decisions in constructing the symbol preference dataset.

\revise{
We run \toolname with different setups to study the effects of individual components. Recall that \toolname outperforms ReSym by 5.6\% on the DIRTY dataset in terms of precision. In this study, we attribute the improvement to different underlying techniques. 
The results are in Fig~\ref{fig:eval-contrib}. 
We can see that all components contribute to the improvements. Specifically, the context-aware fine-tuning paradigm (during training) and the context propagation algorithm (during inference) contribute most to the improvement, indicating the importance of leveraging contextual information.
Moreover, we study how different degrees of contexts sensitivity in context propagation affect the performance of \toolname. The results show that a 5-degree context sensitivity empirically works well. Please see Appendix~\ref{appdx:ctx-sense} for details.
}

\begin{table*}[t]
    \caption{Effectiveness of design decisions in SymPO}
    \label{tab:eval-abl-sympo}
    \centering
    \begin{tabular}{rl ll c ll c ll }
    \toprule
    \multirow{2}{*}{Dataset} & \multirow{2}{*}{\#Pairs} & \multicolumn{2}{c}{Proj. not in train} & & \multicolumn{2}{c}{Proj. in train} && \multicolumn{2}{c}{Overall} \\
    \cmidrule{3-4} \cmidrule{6-7} \cmidrule{9-10}
                 &&   Precision & Recall &              & Precision & Recall  && Precision & Recall\\
    \midrule
        SymPO & 94.3k               & 32.4 & 30.8 && \textbf{42.3} & \textbf{40.6} && \textbf{36.2} & \textbf{34.6} \\ 
    \midrule
SymPO w/o Data Filtering & 232k   & \textbf{32.9}(+0.5) & \textbf{31.1}(+0.4) && 40.3(-1.9) & 38.6(-2.0) && 35.8(-0.5) & 34.0(-0.5) \\
SymPO w/ ground-truth Names & 93.1k & 31.0(-1.5) & 29.5(-1.3) && 40.5(-1.8) & 39.0(-1.6)  &&  34.7(-1.6) & 33.2(-1.4)\\
    \bottomrule
    \end{tabular}
\end{table*}

We further study how each design decision in constructing the symbol preference dataset affects the performance. The results are shown in Table~\ref{tab:eval-abl-sympo}. The default setup (shown in the row \textit{SymPO}) uses the best names predicted by the model as preferred names and uses static feature based heuristics to reduce the size and noise of the  dataset. The second row~(\textit{SymPO w/o Data Filtering}) shows the dataset constructed without the static feature based heuristics. The third row~(\textit{SymPO w/ ground-truth Names}) shows the dataset that uses the ground-truth names as the preferred names.
We can see that the static feature based heuristics reduce the dataset size by around 60\%. And it demonstrates slightly better overall performance than training a model on the whole dataset.
On the other hand, observe that the dataset constructed from ground-truth names results in significantly worse performance than the default setup.

\section{Case Studies}
\label{sec:case-study}

\noindent
{\bf Examples of \toolname's prediction.}
To intuitively demonstrate the effectiveness of \toolname, we show examples of \toolname's prediction that receive each score in the GPT4Evaluator in Fig.~\ref{fig:example-preds} in the Appendix.

\noindent
{\bf Malware reverse engineering.} We use a real-world malware sample~\cite{malware-sample} to illustrate how \toolname helps a security analyst reverse engineer a malware sample. Fig.~\ref{fig:malware-app} shows a code snippet from the studied real-world malware sample. It connects to a command-and-control (C\&C) server, parses the command, and dispatches the commands from the server. In Fig.~\ref{fig:malware-app}a we show the decompiled code generated by IDA~\cite{idapro}. In Fig.~\ref{fig:malware-app}b we show the corresponding code with variables renamed by \toolname. At lines 1--2, the malware receives commands from the server. Lines 3--15 parse the commands, and lines 17--24 dispatch and execute the commands.
We can see that the names predicted by \toolname make the code snippets easier to understand. For example, {\tt i} defined at line 3 is renamed to {\tt tok}. It indicates that the variable stores a \textit{token} of the command. At line 14, the variable {\tt v57} is renamed to {\tt i\_len\_1}. It indicates that the variable stores the length of a sub-component of the variable {\tt i} (now renamed to {\tt tok}). Therefore, it is easier to understand that lines 4--14 split a command into two parts and store them in {\tt dest} and {\tt tok}, respectively~(line 15).
More importantly, \toolname renames {\tt j} at line 17 and {\tt v73} at line 18 to {\tt cmd\_ptr} and {\tt matched\_cmd}, respectively. It reflects that lines 17--24 are dispatching and executing commands from the server. 
This would reveal the suspicious intention of this code snippet.

\noindent
{\bf Binary summarization.} We further study how \toolname helps the binary summarization task. We use \toolname to recover names in a decompiled function. Then we feed the function to ChatGPT and ask ChatGPT to summarize the decompiled function. The study shows that with the predicted variable names, ChatGPT captures more accurate information from the decompiled code. Details are in Appendix~\ref{appdx:bin-sum}.

%% file: related-and-conclusion.tex
\section{Related Work}

\noindent
{\bf Classification-based techniques for recovering symbol names.}
There are existing efforts on reconstructing variable names in stripped binary programs~\cite{dirty,dire,varbert,he2018debin,resym}. DEBIN~\cite{he2018debin} works on BAP-IR~\cite{brumley2011bap} that is more similar to the disassemble code than the decompiled code. It encodes facts in an IR program with probabilistic graph models(PGM) and predicts variable names based on the PGM. DIRTY~\cite{dirty} works on the decompiled code. It leverages a transformer model that interleaves predictions for variable names and variable types. State-of-the-art technique VarBERT~\cite{varbert} also leverages a transformer model working on the decompiled code. Different from DIRTY, which trains the model from scratch on the decompiled code, VarBERT first pre-trains the model on a large corpus of source code and then fine-tunes on the decompiled code. All three techniques formulate the problem as a classification task and thus can hardly predict names unseen in the training dataset. On the other hand, \toolname formulates it as a generative task. It can predict names that rarely appear in the training dataset. 

\noindent
{\bf Generative techniques for recovering symbol names.}
DIRE~\cite{dire} is an early work leveraging a generative model~(i.e., RNN) to solve the renaming problem. Yet it trains the RNN from scratch on a relatively small set of decompiled code and thus underperforms state-of-the-art techniques~\cite{varbert,resym} that can benefit from pre-training efforts on source code.
\revise{
ReSym~\cite{resym} aims to recover names, types, and data structures. It shares a common goal with \toolname on name recovery. It fine-tunes an LLM for renaming variables, and uses program analysis as a post-processing step. In particular, it directly fine-tunes on individual decompiled functions using the ground truth type and name information. It further uses data-flow analysis to propagate type information and data structure fields, and leverages voting to resolve inconsistencies. \toolname goes beyond ReSym by proposing
two unique \textbf{\textit{training}} paradigms~(i.e., Context-aware fine-tuning and SymPO) that more effectively train LLMs on the variable renaming problem. Although the program analysis components in both ReSym and \toolname are adapted from the standard data-flow analysis, they are different in both design and implementation. Specifically, 
the analysis in ReSym focuses on type inference and type checking, whereas  the analysis in \toolname focuses on name propagation, which has a different nature.
}

Besides recovering variable names, another stream of work focuses on recovering function names in decompiled code~\cite{symlm,kim2023transformer}. Their efforts are complementary to ours.~\looseness=-1

\noindent
{\bf Reverse engineering.} Existing efforts~\cite{xu2014autoprobe,zhang2019bda,xu2023pem,jin2023binary,su2024codeart,xu2023lmpa} reverse engineer binary programs to analyze malware~\cite{spensky2016phi,simone23malware}, harden programs~\cite{duck22hardening} and facilitate fuzzing~\cite{fioraldi2020weizz,ntfuzz, scharnowski2022fuzzware}.
Their efforts are complementary with ours, and the results of \toolname can benefit the reverse engineering tasks, as shown in Section~\ref{sec:case-study}.~\looseness=-1

\noindent
{\bf Foundational binary program analysis.} \toolname relies on existing foundational binary analysis techniques~\cite{alves2019function,bauman2018superset,lin2010rewards,brumley2013native} to process binary programs, such as disassembly~\cite{miller2019probabilistic,zhang2021stochfuzz}, type recovery~\cite{shoshitaishvili2016state,lee2011tie,slowinska2011howard,zhang23osprey,zhang2019bda}, and decompilation~\cite{yakdan2015no}. 
State-of-the-art achieves good performance in most cases~\cite{peixda,basque2024ahoy,ye2023d}.~\looseness=-1

\section{Conclusion}

 \revise{ We propose a novel technique that leverages the strengths of generative models to recover meaningful variable names from the decompiled code of fully stripped binary programs. We design context-aware fine-tuning to teach the model how to leverage contextual information, and design symbol preference optimization to mitigate models' biases. 
Our prototype \toolname demonstrates significant improvements on SOTA in challenging setups.}

%% file: appendix.tex
\subsection{How a Reference Model Prevents the SymPO Model from Diverging too much}
\label{appdx:sympo-gradient-analysis}
\begin{figure*}[t]
\input{sympo_gradient}
\end{figure*}

We show the gradient of the SymPO loss in Equations~\ref{eqt:gradient-multi}--~\ref{eqt:gradient-trans}. As shown in Equation~\ref{eqt:gradient-multi}, the gradient is the multiplication of two terms. The second term in the bracket is not affected by the reference model and is straightforward: it enlarges the probability for better names while decreasing the probability for worse names.

On the other hand, the first term constrains the magnitude of the gradient for a given data sample. It can be equally transformed to Equation~\ref{eqt:gradient-trans}. Observe that if the model being optimized already shows significant preference towards the better names compared with the reference model, this term will be close to zero. The updates (to the model weights) introduced by the corresponding data sample will thus be smaller. Therefore, the reference model reduces further optimization on the already learned preference, minimizing the divergence from the reference model.

\subsection{Dataset Preprocessing and Statistics}
\label{appdx:dataset}

\smallskip
\noindent
{\bf Preprocessing following the DIRTY dataset.} We use GHCC to compile C/C++ projects on GitHub created in 2012-2022. Different from DIRTY, (1) we additionally filter out projects with less than 20 stars for quality consideration. (2) We only include executable binary programs in our dataset, precluding intermediate relocatable binary files since the semantics of a relocatable file rely on its symbol table~\cite{compcertelf}, which may be stripped away.  

\smallskip
\noindent
{\bf Rationale of deduplicating binaries by function names.} In our preprocessing pipeline, we conservatively deduplicate binaries by including a binary program only if more than 70\% of its function names are not in the dataset yet. It is common that a project puts the main logic in the shared object (.so) file and keeps other binaries as simple wrapper programs. Take the tool Bibutils~\footnote{\url{https://github.com/biodranik/bibutils}} as an example. The (corresponding source code files of) two binary programs {\tt xml2ris}~\footnote{\url{https://github.com/biodranik/bibutils/blob/master/bin/xml2ris.c}}  and {\tt xml2end}~\footnote{\url{https://github.com/biodranik/bibutils/blob/master/bin/xml2end.c}} are simply two wrapper programs for a shared object {\tt libbibutils.so}. All three binary programs are in the original VarCorpus dataset. However, after including the shared object in the dataset, it is not beneficial to include the two wrapper programs.
As a result, as shown in Table~\ref{tab:dataset-stats}, both our processed datasets are smaller than the original VarCorpus dataset, while their diversity is better than the original VarCorpus dataset.

\begin{table*}[t]
\centering
\caption{Dataset statistics. Each column denotes a dataset. \textit{\#Func} denotes the total number functions in the dataset. \textit{Unique Funcs} denotes the ratio of functions with unique function names. \textit{Unique Name List} denotes the ratio of functions with unique name lists of variables. \textit{\#Vars} denotes the total number of variables, and \textit{Unique Names} denotes the ratio of variables with unique variable names. }
\label{tab:dataset-stats}
\begin{tabular}{rccc}
\toprule
                             & VarCorpus-Ori            & VarCorpus-Our           & DIRTY-Our      \\ \midrule
\#Func                       & \textbf{1,995,847}          & 895,004            & 348,213            \\
Unique Funcs (by name) (\%)   & 46.8               & 81.3               & \textbf{89.4}              \\
Unique Name List (\%)        & 29.6               & \textbf{52.7}              & 40.4               \\ \midrule
\#Vars                       & \textbf{6,126,592}          & 3,363,688          & 1,156,214          \\ 
Unique Names (\%)            & 6.5                & 9.8                & \textbf{12.2}               \\ \bottomrule
\end{tabular}
\end{table*}

\begin{figure*}[t]
\centering
    \begin{subfigure}[t]{.32\linewidth}
    \centering
        \includegraphics[width=\linewidth]{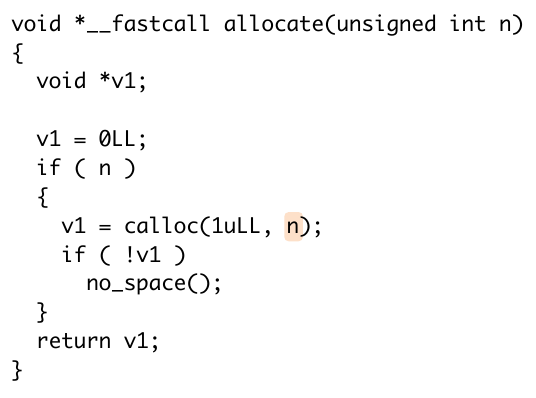}
        \caption{A version of {\tt allocate}}
        \label{fig:appdx:dedup-str1}
    \end{subfigure}    
    ~\hfill
    \begin{subfigure}[t]{.32\linewidth}
    \centering
        \includegraphics[width=.8\linewidth]{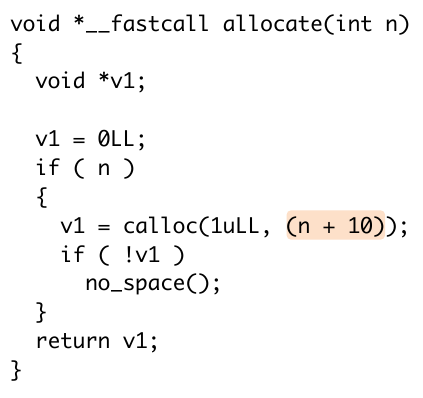}
        \caption{A version almost the same with (a)}
        \label{fig:appdx:dedup-str2}
    \end{subfigure}     
    ~\hfill
    \begin{subfigure}[t]{.32\linewidth}
    \centering
        \includegraphics[width=\linewidth]{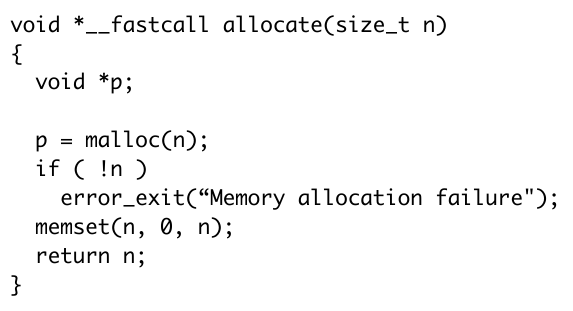}
        \caption{A version different from (a)}
        \label{fig:appdx:dedup-str3}
    \end{subfigure}   
    \caption{Three versions of {\tt allocate}. They demonstrates why checking data leakage with exact string match may still overestimate models' performance. Versions (a) and (b) are almost the same. The difference is highlighted. Version (c) is different from both (a) and (b) because it has different semantics, e.g., setting the allocated memory to zeros. On the other hand, string-similarity can capture the similarity between (a) and (b) while distinguish them with (c). The string similarity scores between (a) and (b), (a) and (c), (b) and (c) are 95, 58, and 58, respectively.}
    \label{fig:appdx:dedpu-str}
\end{figure*}

\begin{table}[t]
    \caption{Performance of models on functions whose highest string similarity score in the training dataset is larger than 90.}
    \label{tab:appdx-overlap-perf}
    \centering
    \begin{tabular}{cc cc}
    \toprule
    Dataset & Model & PR & RC \\            
    \midrule
    \multirow{3}{*}{DIRTY}     & VarBERT     & 50.8 & 50.6 \\ 
    & \toolname$_{\text{Gemma2B}}$          & 59.7 & 58.6 \\
    & \toolname$_{\text{CLM7B}}$          & \textbf{72.3} & \textbf{71.8} \\
    \midrule
    \multirow{2}{*}{VarCorpus} & VarBERT      & 44.4 & 43.7 \\
    & \toolname$_{\text{Gemma2B}}$ & \textbf{56.1} & \textbf{55.1} \\
    \bottomrule
    \end{tabular}
\end{table}

\smallskip
\noindent
{\bf Checking data leakage with string similarity.} We propose to use string similarity, instead of exact string matching, to identify data leakage in the test dataset~(i.e., test functions that are present in the training dataset). 
Previous work~\cite{varbert,dirty,dire} considers two functions as the same only when their normalized strings are exactly the same. 
For example, the VarBERT authors deduplicate the VarCorpus dataset so that all the functions in VarCorpus are not exactly the same. 
Therefore, there is no overlap between the training and test data samples in terms of exact string match. 
However, we observe that considering a sample as data leakage only when there is an exactly matched string in the training data still significantly overestimates the performance of a tested model.

For example, we observe that there are 15,363 functions named {\tt allocate} in the {\it deduplicated} VarCorpus dataset. We show three of them in Fig.~\ref{fig:appdx:dedpu-str} to illustrate the problem. All three versions allocate a chunk of memory and terminate the execution on failure. The two versions in (a) and (b) are only different in the size of allocation. 
They are almost the same function, but cannot be captured by exact string match even after normalization. 
Suppose that version~(a) is in the training dataset. The performance of a model on version~(b) cannot really reflect the generalizability of the model. 
On the other hand, the third function has semantic differences in that it explicitly sets the allocated memory to zero. Therefore, simply considering all functions with the same name as potential data leakage may introduce significant false positives.

We propose to use string similarity~\footnote{\url{https://anhaidgroup.github.io/py_stringmatching/v0.3.x/Ratio.html}} as the metric to conservatively check potential data leakage. The string similarity is calculated based on string edit distance, ranging from 0~(indicating two strings have no overlap) to 100~(indicating two strings are an exact match). Empirically, we consider a test sample as overlapped with the training dataset if the highest string similarity of the sample is larger than 90 to a training data sample.

Table~\ref{tab:appdx-overlap-perf} shows the performance of \toolname and VarBERT on data samples whose highest string similarity to a training data sample is larger than 90. We can see that the performance of all models is substantially better than the performance shown in Table~\ref{tab:eval-performance}. The performance, unfortunately, cannot faithfully reflect the capability of the models on the variable recovery problem.

\subsection{\revise{Correlation between {\tt memset} and {\tt buffer}}}
\label{appdx:corr}
\revise{
We observe that a model is more likely to predict a variable name as {\tt buffer} if the variable is used as the first parameter of {\tt memset}. To quantify our observation, we use Chi-2 test~\footnote{\url{https://docs.scipy.org/doc/scipy/reference/generated/scipy.stats.chisquare.html}} to test the correlation between ``a variable is the first parameter of {\tt memset}'' and ``a variable is predicted the name {\tt buffer}''. The null hypothesis is that the distribution of the two random variables are independent.
As shown in Table~\ref{tab:name-corr}, the results of Chi-2 test reject the null hypothesis with a p-value significantly smaller than 1e-5 (i.e., 1.6e-63), indicating that the two random variables are indeed correlated with a statistical significance. In other words, ``a variable is the first parameter of {\tt memset}'' is indeed correlated with ``a variable is predicted the name {\tt buffer}''.
In comparison, we also run the same test with {\tt memset} and another randomly picked name {\tt file}. The Chi-2 test yields a p-value of 0.22, not supporting the correlation between {\tt memset} and {\tt file}.

}

\begin{table}[t]
    \centering
    \caption{\revise{Correlation between the model's predictions and the corresponding function names. For first two rows, column 1 denotes whether a variable is the first parameter of {\tt memset}, columns 2--3 and 4--5 denote whether a variable is named as {\tt buffer} or {\tt file}, respectively. The last row shows the Chi-2 p-values for {\tt memset} and {\tt buffer}, and {\tt memset} and {\tt file}, respectively. A smaller value denotes higher correlation. }}
    \label{tab:name-corr}
\revise{
\begin{tabular}{c cc c cc}
    \toprule
\multirow{2}{*}{\texttt{memset}} & \multicolumn{2}{c}{\texttt{buffer}} && \multicolumn{2}{c}{\texttt{file}} \\ 
\cmidrule{2-3} \cmidrule{5-6}
                & T & F && T & F \\
\midrule
T   &   19 & 700 &&     2   & 717 \\
F   &   292& 206504 &&  165 & 206631 \\
\midrule
\midrule
$\chi^2$ & \multicolumn{2}{c}{1.6e-63} && \multicolumn{2}{c}{0.22} \\
   \bottomrule
\end{tabular}
}
\end{table}

\input{analysis-and-voting}

\subsection{\revise{Assumptions about Semantics Consistency and Copied Variables}}
\label{appdx:assumption}
\revise{
We have two assumptions about the semantics of variable names. First, we assume that a direct copy between two variables (e.g., {\tt var0 := var1}) implies that the semantics of two variable names are correlated. Second, we assume that the correlation between {\tt var0} and {\tt var1} is not symmetric. Typically, {\tt var0} denotes a broader range of semantics.

To validate the assumptions, we extract 920 pairs of variable names associated with direct copies from the source code of Apache-Httpd~\footnote{\url{https://github.com/apache/httpd}}. For each pair of names, we ask ChatGPT (1) whether the two names have correlated semantics, and (2) whether the names of left-hand-side variables denote a broader range of semantics than the names of right-hand-side variables, and vice versa. As a comparison, we ask ChatGPT the same set of questions on 920 pairs of randomly sampled variable names. The results are shown in Table~\ref{tab:name-corr}.  We can see that name pairs from direct copies have significantly more correlation than random variable pairs. Moreover, for direct copies, we can see that in 50\% of the correlated names, the left hand side name denotes broader semantics while only in 32\% cases, the right hand side denotes broader semantics. The results validate the two assumptions.
}

\begin{table}[t]
    \centering
    \caption{\revise{Correlation between variable name semantics. The first row denotes results for 920 pairs of variables that appear in direct copy operations. The second row shows results for 920 random pairs of variables for reference. Column {\it Corr.} denotes the number of pairs considered as ``correlated semantics''. Columns {\it L $>$ R}, {\it R $>$ L}, and {\it  R $\simeq$ L} denote the left hand side variable name denotes  broader semantics, the right hand side variable name denotes broader semantics, and the semantics of two names are similar, respectively. }}
    \label{tab:name-corr}
    \revise{
\begin{tabular}{r c c c c}
    \toprule
        & Corr. & L $>$ R & R $>$ L & R $\simeq$ L \\
        \midrule
Copied Pairs(920 in total) & 717 & 355 & 226 & 136 \\
Random Pairs(920 in total) & 81 & 35 & 36 & 10 \\
   \bottomrule
\end{tabular}
}
\end{table}

\begin{table}[t]
\centering
\small
\setlength{\tabcolsep}{2pt}
\caption{
Hyper-parameters in \toolname
}
\label{tab:eval:hyper-params}
\centering
\begin{tabular}{rlc}
\toprule
Model & Parameter & Value \\
\midrule
\multirow{4}{*}{Gemma-2B}& batch size & 128 \\
& learning rate scheduler & cosine \\
& learning rate & 5e-5 \\
& warmup steps & 2000 \\
\midrule
\multirow{4}{*}{CodeLlama-7B}& batch size & 64 \\
& learning rate scheduler & cosine \\
& learning rate & 5e-5 \\
& warmup steps & 2000 \\
\midrule
\multirow{4}{*}{CodeLlama-34B}& batch size & 64 \\
& learning rate scheduler & cosine \\
& learning rate & 5e-5 \\
& warmup steps & 500 \\
& LoRA rank & 64 \\
\midrule
\multirow{4}{*}{SymPO(All)}& batch size & 64 \\
& learning rate scheduler & cosine \\
& learning rate & 1e-6 \\
& warmup steps & 100 \\
\bottomrule
\end{tabular}
\end{table}

\begin{figure*}
    \begin{tcolorbox}[]\small
\textbf{System prompt:}

You are an experienced C/C++ reverse engineer.
Please act as an impartial judge and evaluate the quality of names of variables in the given decompiled program.
You will be provided with 
(1) the decompiled code with ground-truth variable names,
(2) the variable names predicted by an AI assistant.

In the evaluation, you should answer the following questions:

\textbf{A. Does the variable name reflect relevant context (domain)? Answer the question in range 5(best) to 1(worst).}

Domain/context describes the high-level program context of the variable. 
It is more of the general high-level domain (e.g., network, memory, CPS, physics, GUI, etc) rather than specific semantics (e.g., len, size, pointer).
\begin{itemize}
    \item For 5, the predicted name and the ground-truth name should describe the same domain/context. 
Or, both the predicted name and the ground-truth name does not explicitly mention a specific domain.
    \item For 4, the domains of the predicted name and the ground-truth name should be similar and relevant, although may not be exactly the same. The predicted name domain may be a superset or subset of the ground truth. 
The predicted domain may be closely related to the ground-truth domain. The predicted name and ground-truth name may be two different perspectives of a same specific domain.
    \item For 3, the predicted name does not explicitly mention a specific context, but the ground-truth name does. 
The predicted name only contains low level operations. 
From the predicted name, one cannot deduce the high-level purpose of the decompiled function/variable.
    \item For 2, the predicted name is slightly misleading. 
The domain for predicted name is different and not relevant to the ground-truth domain. 
However, although misleading, the domain is only implied by the choice of words, and is not explicitly mentioned.
    \item For 1, the predicted name is completely misleading. 
The name is irrelevant to the ground-truth domain, and it is explicitly mentioned in the name.
\end{itemize}

\textbf{B. Does the predicted name reflect relevant semantics? Answer the question in range 1(best) to 5(worst).}

Semantics means the specific high-level meanings denoted by a variable (e.g., len, errmsg, file).
\begin{itemize}
    \item For 5, the semantics of the name should be almost exactly the same to the ground truth.
Or, both the predicted name and the ground-truth name do not have meaningful semantics.
    \item For 4, the semantics of the predicted name are similar to the ground-truth name.
It may be vague, but the overall semantics and purpose is correct.
    \item For 3, the predicted name does not specify meaningful semantics but the ground truth name does. 
It only indicates some low-level operations without high level abstractions.
    \item For 2, the summary specify relevant but inaccurate semantics. 
The semantics specified in the predicted name may be relevant to the ground truth,
but they have significant differences.
    \item For 1, the summary contains irrelevant semantics. 
It denotes a totally different semantics with the ground-truth.
\end{itemize}

You should first briefly summarize the provided decompiled code,
then for each predicted variable name, follow the workflow:

\textbf{Step1: }Output the placeholder variable name you are analyzing, and its ground truth name.

\textbf{Step2: }Explain the ground truth name. (Why it is named like that? What is the high-level context? What is the high-level semantics?)

\textbf{Step3: }Output the predicted name, and explain it.

\textbf{Step4: }Output your score in the format:

\texttt{
\{'var': (ground-truth name here), 'prediction': (predicted name here), 'score': \{'Q-A': [1-5], 'Q-B': [1-5]\}\}
}

Repeat the process for each variable name in the predicted name map.

\textbf{User prompt:}

Decompiled code with ground-truth variable names:
...

Predicted variable names:
...
    \end{tcolorbox}
    \caption{Prompts to GPT4Evaluator}
    \label{fig:prompt-gpt4eval}
\end{figure*}

\subsection{\revise{Reasoning Long Context Remains Challenging for Code Models}}
\label{appdx:long-ctx}

\revise{
Generating code in long coding contexts is a known challenge in the code generation domain\cite{ding2024crosscodeeval,bogomolov2024long}. Although advanced code models~\cite{codellama,guo2024deepseek} achieve decent performance on code generation with relatively short contexts like in the HumanEval dataset~\cite{chen2021evaluating}, their performance drops when the context becomes longer and more complex~\cite{ding2024crosscodeeval,bogomolov2024long}. For example, CodeLlama-70B achieves a pass@1 of 67.8\% on HumanEval~\cite{codellama} yet less than 20\% performance in code generation tasks with significantly longer contexts~\cite{bogomolov2024long}.
}

\subsection{\revise{Effects of Contexts Sensitivity}}
\label{appdx:ctx-sense}

\revise{
We study in the contextual information propagation process, how the degree of context sensitivity affects the results of \toolname. By default, \toolname only propagates names from the direct caller and callee functions of a function (i.e., with 1 degree of context sensitivity). The results are shown in Table.~\ref{tab:context-sense}. We can see that a higher context sensitivity improves the performance of \toolname, yet the improvement becomes marginal when the degree of sensitivity increases from 5 to 10.
}

\begin{table}[t]
    \centering
    \caption{\revise{Effects of context sensitivity}}
    \label{tab:context-sense}
    \revise{
\begin{tabular}{r c c}
    \toprule
Ctx. Degree & PR & RC \\
        \midrule
1 & 35.8 & 35.2 \\
5 & \textbf{36.8} & \textbf{35.5} \\
10 & \textbf{36.8} & \textbf{35.5} \\
   \bottomrule
\end{tabular}
}
\end{table}

\begin{figure}
    \centering
    \includegraphics[width=0.9\linewidth]{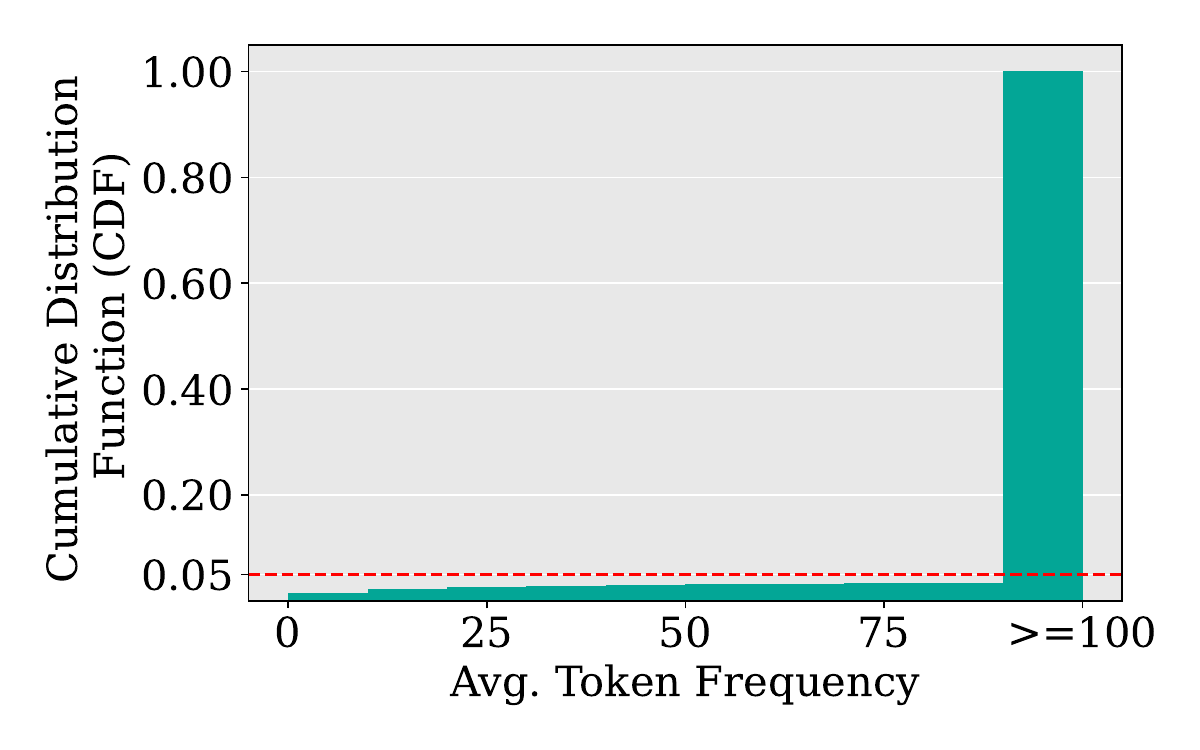}
    \caption{\revise{Token frequency distribution of rarely seen names (i.e., names with training set frequencies less than 10). The x-axis denotes the average {\em token} frequency of a name. That is, we first tokenize the name, and calculate the average frequency of the tokens. The y-axis denotes the cumulative distribution function. We can see that more than 95\% of the rarely seen names can actually be composed from frequently seen tokens.}}
    \label{fig:tok-freq}
\end{figure}

\subsection{Prompts Input to ChatGPT}
\label{appdx:prompt-gpt}

For each model, we start a query with a prompt describing the task and output format, as follows:

\begin{tcolorbox}[]\small
\textbf{Prompt}:
You are a helpful binary program expert. You are helping the user to understand the binary program below.
You will suggest meaningful names for the variables and functions the user asks about.
The asked identifiers are specified in the format of \texttt{Q:[var1,var2,...]}
You will suggest one name for each asked identifier. 
You must output the suggested names in the json format:
\texttt{\{"var1": "suggested\_name1", "var2": "suggested\_name2", ...\}}
\end{tcolorbox}

We evaluate each model with both 0-shot and 3-shot settings. In a 0-shot experiment, we simply follow the prompt with a decompiled function to query. In a 3-shot experiment, we gives each model 3 ``examples'' before each query. In each query, we randomly sample 3 decompiled functions from the training dataset. Then we input both the sampled decompiled functions and the expected output of these functions. After that, we send to the model the query function.

\begin{figure*}[t]
    \centering
\begin{subfigure}{.45\linewidth}
\centering
    \includegraphics[width=\linewidth]{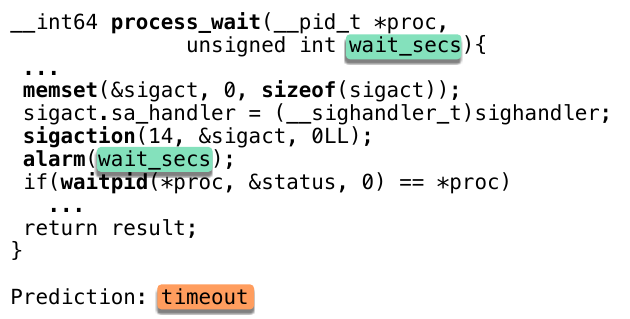}
    \caption{Context:5, Semantics:5. The predicted name has exactly the same semantics and context with the ground-truth name.}    
\end{subfigure}
~\hfill
\begin{subfigure}{.45\linewidth}
\centering
    \includegraphics[width=\linewidth]{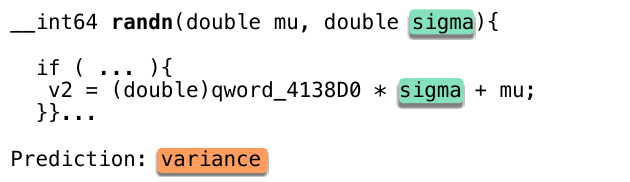}
    \caption{Context:4, Semantics:2. The predicted name has almost the same context as the ground truth (both are related to statistics). However, the semantics is misleading since variance is typically the square of sigma.}    
\end{subfigure}

\begin{subfigure}{.45\linewidth}
\centering
    \includegraphics[width=\linewidth]{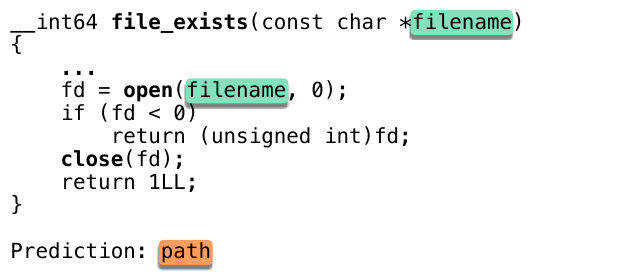}
    \caption{Context:5, Semantics:4. The predicted name is consistent with the program context. However, the semantics of the predicted name does not imply the variable refers to a file. ({\tt path} may also point to a directory.)}    
\end{subfigure}
~\hfill
\begin{subfigure}{.45\linewidth}
\centering
    \includegraphics[width=\linewidth]{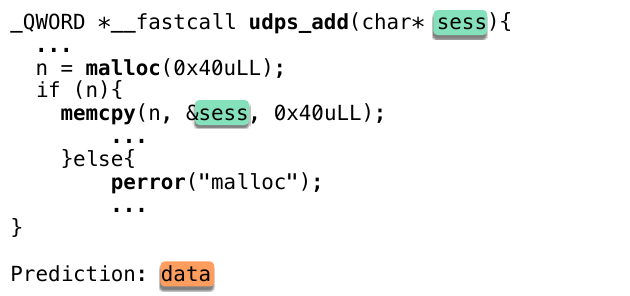}
    \caption{Context:3, Semantics:3. The predicted name does not imply any specific program context, while the ground-truth name {\tt sess} has specific contexts about network. Similarly, the predicted name does not reflect the semantics of this variable, which denotes a ``session''.}    
\end{subfigure}

\begin{subfigure}{.45\linewidth}
\centering
    \includegraphics[width=\linewidth]{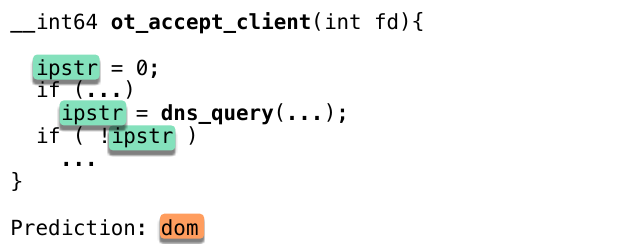}
    \caption{Context:2, Semantics:2. The predicted name {\tt dom} is likely an abbreviation for `domain'. Although it is in general related to network programming, the implied context is not accurate since the ground-truth name implies context about network address. Also, the predicted semantic is misleading since the variable denotes an IP address string, not a domain.}
\end{subfigure}
~\hfill
\begin{subfigure}{.45\linewidth}
\centering
    \includegraphics[width=\linewidth]{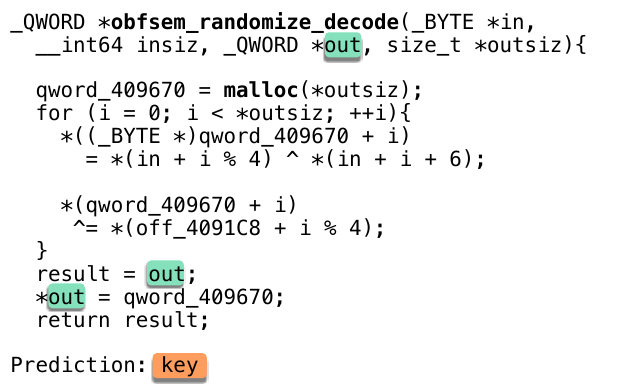}
    \caption{Context:1, Semantics:1. The predicted name is completely misleading and irrelevant to the program context. We speculate \toolname predicts the name {\tt key} because it associates the bit-wise operations in the loop with encryption.}   
\end{subfigure}

    \caption{Examples of \toolname's prediction for each score of GPT4Evaluator. ground-truth names are highlighted in green and predicted names are highlighted in orange.}
    \label{fig:example-preds}
\end{figure*}

\subsection{Case Study: Binary Summarization}

\label{appdx:bin-sum}

\begin{figure}
    \centering
    \begin{subfigure}[t]{\linewidth}
    \centering
        \includegraphics[width=\linewidth]{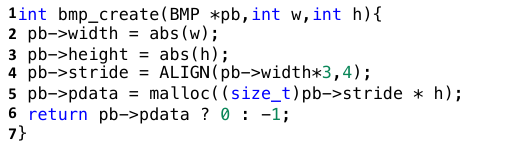}
        \caption{Source code}
        \label{fig:binsum-src}
    \end{subfigure}
    \begin{subfigure}[t]{\linewidth}
    \centering
        \includegraphics[width=\linewidth]{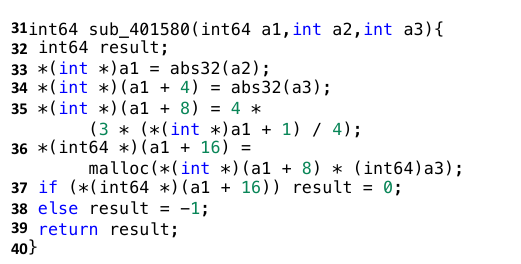}
        \caption{Decompiled code}
        \label{fig:binsum-decomp}
    \end{subfigure}    
    \begin{subfigure}[t]{\linewidth}
    \centering
        \includegraphics[width=\linewidth]{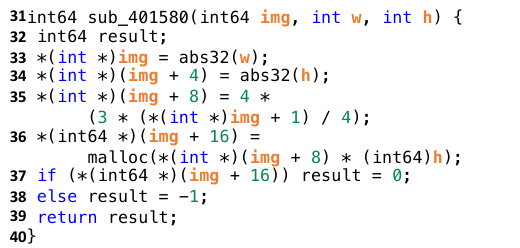}
        \caption{Renamed code (generated names highlighted in orange)}
        \label{fig:binsum-rename}
    \end{subfigure}
    \caption{Binary summarization}
    \label{fig:binsum-app}
\end{figure}

Binary summarization helps a reverse engineer understand a binary function by generating natural language summary for the function.~\cite{al2023extending, jin2023binary}. Existing work show that the state-of-the-art code models still have gaps understanding decompiled code due to the lack of symbols~\cite{jin2023binary}. We show an example in Fig.~\ref{fig:binsum-app} to illustrate how names generated by \toolname benefit the binary summarization downstream task. We show the source code, the decompiled code, and the renamed code in Fig.~\ref{fig:binsum-src},~\ref{fig:binsum-decomp},~and~\ref{fig:binsum-rename}, respectively. The code snippet initializes a BMP image structure. It calculates the absolute values of the width and height (lines 2--3 in Fig.~\ref{fig:binsum-src}), calculates the row size of the image (line 4), allocates a chunk of memory for the image (line 5), and returns a value indicating whether the initialization succeeds (line 6).

We input the three versions to ChatGPT and show part of its response below. We use numbers to label the sentences and use red texts to emphasize extra information obtained from the renamed code (compared to the decompiled code). The full response from ChatGPT are shown as follows:

\begin{tcolorbox}[]\small
\textbf{Source code}:
This function, {\tt bmp\_create}, \circled{1}\textit{initializes a BMP structure} with specified width {\tt w} and height {\tt h}. \circled{2}It sets the \textit{width and height} of the BMP structure to the \textit{absolute values} of {\tt w} and {\tt h}, \circled{3} calculates the {\tt stride} (which is the \textit{aligned width considering a 4-byte boundary}), and \circled{4} \textit{allocates memory} for the pixel data ({\tt pdata}). The function \circled{5} \textit{returns 0 if memory allocation is successful, and -1 otherwise}.
\end{tcolorbox}

\begin{tcolorbox}[]\small
\textbf{Decompiled code}:
This function, {\tt sub\_401580}, \circled{1} takes three arguments and performs a series of operations on them. \circled{2} It calculates the absolute values of a2 and a3 and stores them at specific memory locations relative to a1. \circled{3} It then calculates a value based on these absolute values and stores it at another location. \circled{4} Finally, the function attempts to allocate memory proportional to the calculated value and a3. \circled{5}  If the memory allocation is successful, the function returns 0; otherwise, it returns -1. This suggests the function is used for setting up a data structure with dynamically allocated memory.
\end{tcolorbox}

\begin{tcolorbox}[]\small
\textbf{Renamed code}:
This function, {\tt sub\_401580}, \circled{1} initializes {\color{red} an image structure}. It takes three parameters: a pointer to the image structure img, and two integers {\color{red} w and h for width and height}. \circled{2} The width and height are stored as absolute values in the image structure. \circled{3} It then calculates and stores {\color{red} the row size}, aligning it to {\color{red} a 4-byte boundary}. \circled{4} Memory is allocated for {\color{red} the image data}, the size of which is based on the calculated row size and height. \circled{5} The function returns 0 if the memory allocation is successful, or -1 (indicating an error) if the allocation fails.
\end{tcolorbox}

We can see that the description for the renamed code is closer to that of the source code. It is more relevant to the context (e.g., in \circled{1} ``initializes an image structure'' vs ``setting up a data structure'') and is more accurate at the details (e.g., in \circled{3}, the description for the decompiled version misses the ``4-byte boundary alignment'').

%% file: sympo_gradient.tex
\begin{align}
\label{eqt:gradient-multi}
\nabla_{\Theta}\mathcal{L}_{\mathrm{SymPO}}(\Theta, \Theta_{\mathrm{ctx}})  \Coloneqq &
 -\beta\mathbb{E}_{(q,b,w)\sim D_{\mathrm{prf}}}
\Bigg[
\delta(q,b,w,\Theta, \Theta_{\mathrm{ctx}})
\Big[
\underbrace{
\nabla_{\Theta} \log \mathbf{P}(b|q;\Theta)
}_{
\text{
\begin{minipage}{.15\linewidth}
\centering
    increase preference towards \textbf{\textit{better}} symbols
\end{minipage}
}
}
-
\underbrace{
\nabla_{\Theta} \log \mathbf{P}(w|q;\Theta)
}_{
\text{
\begin{minipage}{.15\linewidth}
\centering
    decrease preference towards \textbf{\textit{worse}} symbols
\end{minipage}
}
}
\Big]
\Bigg]
\\
\label{eqt:gradient-delta}
\delta(q,b,w,\Theta, \Theta_{\mathrm{ctx}})  \Coloneqq &
\sigma\Big(
\beta \log \frac{\mathbf{P}(w|q;\Theta)}{\mathbf{P}(w|q;\Theta_{\mathrm{ctx}})}
-
\beta \log \frac{\mathbf{P}(b|q;\Theta)}{\mathbf{P}(b|q;\Theta_{\mathrm{ctx}})}
\Big)
\\
\label{eqt:gradient-trans}
= &
\sigma\Bigg(
\beta \Big( \log \frac{\mathbf{P}(w|q;\Theta)}{\mathbf{P}(b|q;\Theta)}
-
\log \frac{\mathbf{P}(w|q;\Theta_{\mathrm{ctx}})}{\mathbf{P}(b|q;\Theta_{\mathrm{ctx}})}
\Big)
\Bigg)
\end{align}

%% file: analysis-and-voting.tex
\subsection{\revise{Program Analysis and Semantics Voting at the Inference Stage}}
\label{appdx:pa-and-voting}

This section discusses the name validation algorithm~(Step 3 in Fig.~\ref{fig:wf-infer}) that
leverages program analysis to aggregate the names predicted in different rounds under individual contexts.
The insight is
that names correlated through data flow ought to have a certain level of consistency (in terms of their natural language semantics), although they may not be identical.
For example, a variable named {\tt fout} may be passed as an argument to a parameter named {\tt stream}, but is less likely to be assigned to a variable named {\tt size}. 

The name validation process in \toolname first extracts correlated name candidates for a variable and then selects the candidate with the maximum level of consistency. That is, our goal is to achieve a minimal total semantics distance for all 
correlated names.
We formally define the name validation process in Algorithm~\ref{alg:name-valid}.
The algorithm takes as input a binary program, an initial name map from a variable to its initially selected name,
and a map from each variable to a list of its candidate names.
It outputs an updated name map from each variable to its (updated) name. At the beginning of the name validation process, the initial name map is constructed as a map from a variable to the top-1 predicted name of the generative model.
The name validation algorithm consists of a loop. In each iteration, it collects additional names for each variable by inheriting names from other variables that have {\em direct} data flow with the variable (line~\ref{alg:line:ext-cor} and details explained later), selects the best name of a variable by \textit{semantics voting} (line~\ref{alg:line:sem-vot}), which will be explained later in the section, and updates the name map. It may take multiple iterations to  propagate names to places that are multiple data-flow edges away. The algorithm terminates when no variable is updated or until a predefined budget is reached.

\begin{algorithm}[t]
\caption{Name Validation}
\small
\label{alg:name-valid}
\KwIn{$B$: a binary program}
\KwIn{$N_0$: $\textit{id} \rightarrow \textit{id} \rightarrow \textit{str}$, a map from a variable to an initially selected name}
\KwIn{$\hat{N}$: $\textit{id} \rightarrow \textit{id} \rightarrow \textit{list}\ \textit{str}$, a map of name candidates}
\KwOut{$N$: $\textit{id} \rightarrow \textit{id} \rightarrow \textit{str}$, a map from a variable to the name selected by the algorithm}
\BlankLine

\textit{update} $\leftarrow \textit{True}$

$N_{current} \leftarrow N_0$

\While{update}{
    $N_{prev} \leftarrow N_{current}$ 
    
    $\hat{N}' \leftarrow$  \textit{correlated\_names}($B$, $N_{current}$) \label{alg:line:ext-cor}

    $N_{current} \leftarrow $ \textit{semantics\_vote}($\hat{N}$, $\hat{N}'$)\label{alg:line:sem-vot}

    \textit{update} $\leftarrow N_{prev} \neq N_{current}$
}
\Return{$N_{current}$}
\end{algorithm}

\noindent
{\bf Collecting correlated name by data-flow.} 
If a variable is directly copied to another variable, called {\em a direct use of the variable}, their names should be semantically consistent, analogous to type consistency. In this step, we propagate names across such direct uses to populate the candidate set and enable inconsistency suppression. Note that such propagation is not performed for composite operations. For example, the name of  a right-hand-side variable involved in a binary operation may not be semantically consistent with the name of either left-hand-side variable.

We model the correlation extraction process as finding solutions to program constraints. The key constraint rules are shown in Fig.~\ref{fig:infer-rules}. The analysis takes as input a binary program and the corresponding name map. Its outputs are a correlated variable map $\pi$ that maps from a variable (referred by a function id and a variable id) to a list of correlated names. 
For each variable, the auxiliary data structure $\sigma$ maintains the origin of each correlated name to prevent duplication. Note that if an origin variable has many data-flow paths to another variable, its name may get propagated multiple times and have an in-appropriate weight in the later voting step.
Specifically, an element in $\sigma$ is a triple, noted as $(fid, vid, name)$, where $fid$ and $vid$ refers to the origin variable of a correlated name $name$.

A rule $\frac{A\ B}{C}$ is interpreted as follow: when $A$ and $B$ are satisfied, $C$ is satisfied. The notation $env \vdash Stmt: env'$ is interpreted as given an environment $env$, the environment $env'$ satisfies the constraints introduced by $Stmt$. The two special rules \textbf{Init} and \textbf{Out} are evaluated only once at the beginning and ending of the analysis, respectively. \textbf{Init} initializes the correlated name set of a variable to its initially selected name. The rule \textbf{Out} converts the correlated name set to the correlated name list for all variables.

The rule \textbf{Assign} depicts the constraint introduced by an assignment statement from $id_1$ to $id_0$. The rule requires all the correlated names of $id_1$ to be in the correlated name set of $id_0$ as well.
We assume the name of (the destination variable) $id_0$ to be more general than the name of (the source variable) $id_1$. Therefore, the names correlated to $id_1$ should also have correlation with $id_0$. %
For example, assume an assignment statement {\tt ptr=msg}. The name {\tt ptr} is more general than the name {\tt msg}. Thus a correlated name of {\tt msg} (e.g., {\tt buffer}) is likely to be a correlated name of {\tt ptr} as well.
Moreover, as depicted by the rule \textbf{Assign-R}, we propagate the selected name of $id_0$ to $id_1$ because the denotation 
of $id_0$ and $id_1$ may be similar. However, we do not propagate the correlated names of $id_0$ to $id_1$ because not all the names correlated to $id_0$ are necessarily correlated to $id_1$, assuming the name of $id_0$ denotes a broader range of semantics than the name of $id_1$. \revise{We quantify our assumptions about variable name semantics in Appendix~\ref{appdx:assumption}.}

The intuition of rules \textbf{Call} and \textbf{Ret} are the same. A function call would have implicit assignments between the arguments and the parameters. And the function return would have an implicit assignment between the return value and the variable storing calling results in the caller function. The dual rules \textbf{Call-R} and \textbf{Ret-R} can be interpreted similarly.

\newcommand{\progstate}[2]{\ensuremath{\langle #1, #2 \rangle}\xspace}
\newcommand{\update}[3]{\ensuremath{#1\big[#2\rightsquigarrow#3\big]}\xspace}
\newcommand{\concat}{\ensuremath{+\!\!\!+}\xspace}
\begin{figure*}[t]
    \centering
\begin{minipage}{.8\linewidth}
\begin{mdframed}
\small
    \textbf{Input:} $B:\mathcal{B},\ N_{in}:\mathcal{N}$\quad
    \textbf{Output:} $\pi: \textit{id} \rightarrow \textit{id} \rightarrow \textit{list}\ \textit{str}$ 
    
    \textbf{Auxiliary Data:} $\sigma: \textit{id} \rightarrow \textit{id} \rightarrow \textit{set}\ (\textit{id} \times \textit{id} \times \textit{str})$ \quad
    \textbf{State Configuration:} \progstate{\pi}{\sigma}

    We use $f$ to denote the function that the analyzed statement belongs to.
\end{mdframed}
\end{minipage}
\begin{mathpar}
\small
\inferrule{
\sigma_0[fid][vid] = n \\ (fid, vid, n) \in N_{in}
}{
\progstate{\cdot}{\cdot} \vdash: \progstate{\cdot}{\sigma_0}
}
{\textbf{Init
}}

\quad
\inferrule{
\pi[fid][vid] = \big[ n| (\cdot, \cdot, n) \in \sigma[fid][vid] \big] \\ (fid, vid) \in \sigma
}{
\progstate{\cdot}{\sigma} \vdash \textit{Done}: \progstate{\pi}{\sigma}
}
{\textbf{Out}}
\\
\inferrule{
\sigma' =  \update{\sigma}{f.\fieldemph{fid},id_0}{\sigma[f.\fieldemph{fid}][id_0]\cup\sigma[f.\fieldemph{fid}][id_1]}
}{
\progstate{\cdot}{\sigma} \vdash id_0 \coloneqq id_1 :\progstate{\cdot}{\sigma'}
}
{\textbf{Assign}}
\quad
\inferrule{
n_1 = N_{in}[f.\fieldemph{fid}][id_0]\\\\
\sigma' =  \update{\sigma}{f.\fieldemph{fid},id_1}{\sigma[f.\fieldemph{fid}][id_1]\cup\{(f.\fieldemph{fid}, id_0, n_1)\}}
}{
\progstate{\cdot}{\sigma} \vdash id_0 \coloneqq id_1 :\progstate{\cdot}{\sigma'}
}
{\textbf{Assign-R}}
\\
\inferrule{
f_1\in B.\textbf{\textit{funcs}}\\
id_0 = args[i] \\ 
p_0 = r[i] \\
r, \_ = f_1.\fieldemph{body}\\\\
\sigma' = \update{\sigma}{f_1.\fieldemph{fid}, p_0}{\sigma[f_1.\fieldemph{fid}][p_0]\cup\sigma[f.\fieldemph{fid}][id_0]}
}{
\progstate{\cdot}{\sigma} \vdash f_1.\fieldemph{fid}(args) :\progstate{\cdot}{\sigma'}
}
{\textbf{Call}}
\quad
\inferrule{
f_1\in B.\textbf{\textit{funcs}}\\
id_1\coloneqq f.\fieldemph{fid}(...) \in f_1.\textbf{\textit{body}}\\\\ 
\sigma' = \update{\sigma}{f_1.\fieldemph{fid}, id_1}{\sigma[f_1.\fieldemph{fid}][id_1]\cup\sigma[f.\fieldemph{fid}][id_0]}
}{
\progstate{\cdot}{\sigma} \vdash \langkw{return } id_0 :\progstate{\cdot}{\sigma'}
}
{\textbf{Ret}}
\\
\inferrule{
f_1\in B.\textbf{\textit{funcs}}\\
id_0 = args[i]\\ 
p_0 = r[i] \\
r,_ = f_1.\fieldemph{body}
\\\\
n_1 = N_{in}[f_1.\fieldemph{fid}][p_0]\\
\sigma' = \update{\sigma}{f.\fieldemph{fid}, id_0}{\{(f_1.\fieldemph{fid}, p_0, n_1)\}\cup\sigma[f.\fieldemph{fid}][id_0]}
}{
\progstate{\cdot}{\sigma} \vdash f_1.\fieldemph{fid}(args) :\progstate{\cdot}{\sigma'}
}
{\textbf{Call-R}}
\\
\inferrule{
f_1\in B.\textbf{\textit{funcs}}\\
id_1\coloneqq f.\fieldemph{fid}(...) \in f_1.\textbf{\textit{body}}\\\\ 
n_1 = N_{in}[f_1.\fieldemph{fid}][id_1]\\
\sigma' = \update{\sigma}{f.\fieldemph{fid}, id_0}{\sigma[f.\fieldemph{fid}][id_0]\cup\{(f_1.\fieldemph{fid}, id_1, n_1)\}}
}{
\progstate{\cdot}{\sigma} \vdash \langkw{return } id_0 :\progstate{\cdot}{\sigma'}
}
{\textbf{Ret-R}}
\\
\inferrule{
\progstate{\cdot}{\sigma} \vdash \mathcal{S}_1 : \progstate{\cdot}{\sigma_1} \\
\progstate{\cdot}{\sigma_1} \vdash \mathcal{S}_2 : \progstate{\cdot}{\sigma_2} \\
}{
\progstate{\cdot}{\sigma} \vdash \mathcal{S}_1;\mathcal{S}_2 :\progstate{\cdot}{\sigma_2}
}
{\textbf{Step}}
\quad
\inferrule{
\progstate{\cdot}{\sigma} \vdash S:\progstate{\cdot}{\sigma}
}{
\progstate{\cdot}{\sigma} \vdash \langkw{while}(\mathcal{E})\{\mathcal{S}\}  :\progstate{\cdot}{\sigma}
}
{\textbf{While}}
\\
\inferrule{
\progstate{\cdot}{\sigma} \vdash \mathcal{S}_1 : \progstate{\cdot}{\sigma_1} \\
\progstate{\cdot}{\sigma} \vdash \mathcal{S}_2 : \progstate{\cdot}{\sigma_2} \\
\big((fid,vid) \in \sigma_1 \lor (fid,vid)\in \sigma_2\big)\\\\
\sigma_3[fid][vid] = \sigma_1[fid][vid] \cup \sigma_2[fid][vid]
}{
\progstate{\cdot}{\sigma} \vdash \langkw{if}(\mathcal{E})\{\mathcal{S}_1\}\langkw{else}\{\mathcal{S}_2\}  :\progstate{\cdot}{\sigma_3}
}
{\textbf{If}}
\quad 
\inferrule{
}{
\progstate{\cdot}{\sigma} \vdash \textit{Other Stmts}: \progstate{\cdot}{\sigma}
}
{\textbf{Default}}
\end{mathpar}
    \caption{Correlation Extraction Rules 
    }
    \label{fig:infer-rules}
    \vspace{-5pt}
\end{figure*}

\noindent
{\bf Semantics voting.} Semantics voting takes as input the candidate name map $\Hat{N}$ (produced by the model) and the correlated name map $\Hat{N}'$. For each variable, it picks the candidate name that is most similar to all correlated names and other candidate names. It is hard to directly compare the semantics similarity of two strings. Therefore, our algorithm encodes all correlated names and all candidates names to their embeddings, and measures the similarity between two names by calculating the cosine similarity.
Formally, the semantics voting process for a given variable is shown as follows:
\begin{align}
\small
\begin{aligned}
    Candi \coloneqq \Hat{N}[fid][vid]\quad Corr \coloneqq \Hat{N}'[fid][vid] \nonumber \\
    \forall fid\ vid, \argmax_{n\in Candi} \sum_{m \in  [Corr; Candi]} \langle \mathbf{e}_{n}, \mathbf{e}_{m} \rangle
\end{aligned}
\end{align}
where $\langle \cdot, \cdot \rangle$ denotes cosine similarity between two embeddings and $[\cdot ; \cdot]$ denotes list concatenation.

%% file: artifact.tex
\clearpage\newpage
{
\section*{Artifact Appendix}
}
\setcounter{subsection}{0}

\subsection{Description \& Requirements}

Our paper, \toolname, proposes a large language model based reverse engineering technique that recovers variable names from stripped binaries.
Specifically, it takes as input the decompiled code of a stripped binary program.
Decompiled binary program has a syntax that is similar to the C programming language.
However, it does not contain meaningful variable names.
The names in the variables are just placeholders like \texttt{var\_1}, \texttt{var\_2}, etc.
\toolname aims to recover meaningful variable names for those variables.

{\em NOTE: The artifact evaluation is on a prior version~(i.e., the version before the major revision) of our paper.}

\subsubsection{How to access}
Our artifact contains detailed explanations and step-by-step instructions for running the experiments. Here is the DOI of our artifact: \url{https://zenodo.org/records/14220042}, and the corresponding GitHub repository: \url{https://github.com/XZ-X/gennm-ndss-ae}. The preprocessed datasets, model checkpoints, and the intermediate results are uploaded at Zenodo \footnote{\url{https://zenodo.org/records/14287032}}.

\subsubsection{Hardware dependencies}
At least 16GB RAM, a GPU with at least 24GB VRAM.

\subsubsection{Software dependencies} 
Ubuntu 20.04 or later. Anaconda (a Python package manager).

\subsubsection{Benchmarks} 
We provided all the necessary data in the aforementioned Zenodo link. We use two datasets. One dataset is collected from GitHub following a similar setup of a previous work, DIRTY. The other dataset is reused from a previous work, VarBERT.

\subsection{Artifact Installation \& Configuration}

Please download the Zenodo package containing data, model checkpoints, and intermediate results, and unzip the data package under the root directory of the artifact repository.

Then, create a new environment and install the dependencies by running the following commands:

\begin{tcolorbox}[]
\tt\small
\$ conda create -n gennm-artifact python=3.10

\$ conda activate gennm-artifact

\$ pip install -r requirements.txt
\end{tcolorbox}

The \texttt{README.md} file in the code repository contains a brief introduction to the file structures of the artifact.

\subsection{Major Claims}

\begin{itemize}
    \item (C1): \toolname outperforms the state-of-the-art technique, VarBERT, in terms of precision and recall. This is proven by experiments (E1) and (E6) whose results are illustrated by Table~1 in the paper.
    \item (C2): \toolname can generalize to different decompilers and different optimization levels. This is proven by experiment (E2)  whose results are illustrated by Fig~13 in the paper.
    \item (C3): \toolname has better generalizability on names that are rarely seen in the training dataset. This is proven by experiment (E3) whose results are illustrated by Fig~12 in the paper.
    \item (C4): \toolname outperforms VarBERT when evaluated by a GPT-Evaluator that mimics how a human would perceive the results. This is proven by experiment (E4) whose results are illustrated by Fig~11 in the paper.
    \item (C5): \toolname outperforms state-of-the-art black-box LLMs. This is proven by experiment (E5) whose results are illustrated by Table~2 in the paper.    
\end{itemize}

\subsection{Evaluation}

This section contains the same instructions as in the {\tt README.md} file of our code repository.

\subsubsection{Experiment (E1) [10 human-minutes + 10 compute-minutes]}

This experiment reproduce the results in Table 1. 
\textbf{Please run the following command:}

\begin{tcolorbox}[]
\tt\small
scripts/eval\_1\_compare\_dirty.sh
\end{tcolorbox}

This script will first load the output of both \toolname-2B and VarBERT.
Then it calculates the average precision and recall for both \toolname-2B and VarBERT.
The results should be in the following format:

\begin{tcolorbox}[]
\tt\small
proj\_IT     gennm\_pr  gennm\_rc  varbert\_pr  varbert\_rc
                                                                  
False                 0.305068      0.287518           0.235534        0.217368
True                  0.416923      0.395864           0.313501        0.296283
\end{tcolorbox}

Each row denotes whether the function is in a project that is overlapped with the training dataset.
For example, the first row denotes the functions that are not in the training dataset.
{\tt gennm\_pr} and {\tt gennm\_rc} corresponding to the row {\tt DIRTY-GenNm-CG-2B} and {\tt Proj. NIT} columns of Table~1.
{\tt varbert\_pr} and {\tt varbert\_recall} corresponding to the row {\tt DIRTY-VarBERT} and {\tt Proj. NIT} columns of Table~1.

Similarly, the second row denotes the functions that are in the training dataset.
It corresponds to the columns {\tt Proj. IT} for rows {\tt DIRTY-GenNm-CG-2B} and {\tt DIRTY-VarBERT} of Table~1, respectively.

\noindent
\textbf{Please run the following command }to compute the performance for GenNm-CLM-7B: 
\begin{tcolorbox}[]
\tt\small
scripts/eval\_2\_compare\_dirty-7b.sh
\end{tcolorbox}

Note that the results of VarBERT will be printed again. They denote the same results as the previous script.
There might be minor differences (less than 0.005) than the results in the paper due to the randomness of the inference process.

\noindent
\textbf{Please run the following script} to reproduce the rows for VarCorpus in Table 1:
\begin{tcolorbox}[]
\tt\small
scripts/eval\_3\_compare\_ida-O0.sh
\end{tcolorbox}
The results can be interpreted in the same way as the previous scripts.

\subsubsection{Experiment (E2) [5 human-minutes + 10 compute-minutes]}

This experiment aims to reproduce Fig.~13. Fig.~13 shows that \toolname outperforms VarBERT in different decompilers and optimization levels.
In Fig.~13, the x-axis labels {\tt In-PR} and {\tt In-RC} denote the precision and recall for {\tt proj\_in\_train=True}, 
and {\tt Not-PR} and {\tt Not-RC} denote the precision and recall for {\tt proj\_in\_train=False}.

\noindent
\textbf{Please run the following script }to compute the results for {\tt IDA-O3} (the left sub-fig of Fig.~13):

\begin{tcolorbox}[]
\tt\small
scripts/eval\_4\_compare\_ida-O3.sh
\end{tcolorbox}

\noindent
\textbf{Please run the following script} to compute the results for {\tt Ghidra-O0} (the right sub-fig) of Fig.~13.

\begin{tcolorbox}[]
\tt\small
scripts/eval\_5\_compare\_ghidra-O0.sh
\end{tcolorbox}

The outputs of both scripts have the same format as the scripts for Table 1.

\subsubsection{Experiment (E3) [5 human-minutes + 5 compute-minutes]}

This experiment aims to reproduce Fig.~12 which shows that GenNm has better performance than the baseline for variables that are rarely seen during training.

\noindent
{\bf Please run the following script} to reproduce Fig.~12:
\begin{tcolorbox}[]
\tt\small
scripts/eval\_6\_frequency.sh
\end{tcolorbox}

It first computes the frequency of names in the training dataset, and aggregates the performance of both GenNm and the baseline by the name frequencies.
The output should look similar to the following:
\begin{tcolorbox}[]
\tt\small
Freq 0: GenNm-PR: 0.226..., VarBERT PR: 0.084...
...
\end{tcolorbox}
The difference should be minor (less than 0.005) compared to the expected results.

\subsubsection{Experiment (E4) [5 human-minutes + 10 compute-minutes]}

This experiment aims to reproduce Fig.~11. Fig.11 shows the performance of both GenNm and the baseline evaluated by GPT4.
We ask ChatGPT to evaluate each name by two questions, that is, Context Relevance (noted as Q-A in our scripts) and
Semantics Relevance (noted as Q-B in our scripts).

\noindent
{\bf Please run the following script } to reproduce the results:
\begin{tcolorbox}[]
\tt\small
scripts/eval\_7\_gpt4eval.sh
\end{tcolorbox}

It outputs the distribution of the score for each question. The output looks similar to the following:
\begin{tcolorbox}[]
\tt\small
gennm\_qa\_score\\
1    189\\
2    139\\
...
\end{tcolorbox}
For example, {\tt gennm\_qa\_score} denotes the aggregated scores of variables names generated by \toolname. {\tt 1 189} denotes there are 189 names obtained the score {\tt 1}. 

\subsubsection{Experiment (E5) [5 human-minutes + 10 compute-minutes]}

This experiment reproduces the results of Table~2, which comapres the performance of \toolname with black-box LLMs.

\noindent
{\bf Please run the following command} to reproduce the results:
\begin{tcolorbox}[]
\tt\small
scripts/eval\_8\_compare\_blackbox\_llm.sh
\end{tcolorbox}

\subsubsection{Experiment (E6) [5 human-minutes + 240 compute-minutes]}

This experiment is \textit{optional}. For reviewers who have access to a GPU, we provide the following scripts to run \toolname on a small subset of DIRTY and a small subset of VarCorpus. For both data-subsets, we can observe \toolname outperforms the baseline VarBERT.

\noindent
{\bf Please run the following command} to run \toolname on the subset of DIRTY:
\begin{tcolorbox}[]
\tt\small
scripts/infer\_1\_generation.sh 
\end{tcolorbox}

\noindent
{\bf Please run the following command} to run \toolname on the subset of VarCorpus:
\begin{tcolorbox}[]
\tt\small
scripts/infer\_2\_generation-ida-O0.sh
\end{tcolorbox}
It takes less than 2 hour to finish \textit{each} command on a machine with one A6000 GPU, respectively. The expected results are that the performance of \toolname is consistently better than VarBERT.